\documentclass[aps,twocolumn,pra,amsmath,amssymb,superscriptaddress,floats]{revtex4}

\usepackage{amsmath}	%allows the use of mathematical symbols
\usepackage{gensymb}
\usepackage{hyperref}
\hypersetup{colorlinks=true}
\usepackage{upgreek}
\usepackage{graphics}
\usepackage{hyperref}
\usepackage{epsfig}
\usepackage{color}
\usepackage{bm}
\usepackage{float}
\usepackage{ulem}           % \sout
\usepackage{dsfont}         % \mathds
\usepackage{wasysym}

\usepackage{accents}

\usepackage{graphicx}

\graphicspath{{./}{./plots/}}

\usepackage{CJK}
\usepackage{url}
\usepackage{multirow}
\newcommand{\ii}{\imath}   % imag unit

\begin{document}

\title{Interplay of magnetic field and trigonal distortion in honeycomb $\Gamma$ model:\\ Occurrence of a spin-flop phase}
\author{Qiang Luo}
%\email[]{qluo@physics.utoronto.ca}
\affiliation{Department of Physics, University of Toronto, Toronto, Ontario M5S 1A7, Canada}
\affiliation{College of Physics, Nanjing University of Aeronautics and Astronautics, Nanjing, 211106, China}

\author{Hae-Young Kee}
\email[]{hykee@physics.utoronto.ca}
\affiliation{Department of Physics, University of Toronto, Toronto, Ontario M5S 1A7, Canada}
\affiliation{Canadian Institute for Advanced Research, Toronto, Ontario, M5G 1Z8, Canada}
\date{\today}

\begin{abstract}
    In candidate Kitaev materials, the off-diagonal $\Gamma$ and $\Gamma'$ interactions are identified to come from the spin-orbit coupling and trigonal distortion, respectively.
    They have generated intense research efforts because of their intimate relation to the field-induced magnetically disordered state reported in $\alpha$-RuCl$_3$.
    Theoretically, while a plethora of field-induced phases has been proposed in the honeycomb lattice,
    a stable intermediate phase that can survive in a wide parameter region regardless of the underlying phases is still lacking.
    Here we focus on the interplay of an out-of-plane magnetic field and a symmetry-allowed $\Gamma'$ term due to trigonal distortion
    in the dominant antiferromagnetic $\Gamma$ region.
    By using multifaceted approaches ranging from classical Monte Carlo and semiclassical spin-wave theory to density-matrix renormalization group,
    we identify an intriguing spin-flop phase in the presence of magnetic field and antiferromagnetic $\Gamma'$ interaction,
    before it eventually enters into a fully polarized state.
    As the $\Gamma'$ interaction approaches the size of $\Gamma$ one,
    the $\Gamma$-$\Gamma'$ model maps to the easy-axis XXZ antiferromagnet,
    where the spin-flop phase can be understood as a superfluid phase in the extended Bose-Hubbard model.
    Our finding thus demonstrates an exciting path from the honeycomb $\Gamma$ model towards a $U(1)$-symmetric XXZ antiferromagnet in a magnetic field.
\end{abstract}

%\keywords{}

\maketitle

\section{Introduction}\label{SEC:INTRO}
In the pursuit of exotic quantum ground states such as quantum spin liquid (QSL),
a large family of spin-orbit coupled effective spin-$1/2$ Mott insulators on a honeycomb lattice
has been the focus of massive research efforts~(for reviews, see Refs.~\cite{RauLeeKee2016,TakagiTJ2019}).
This interest is triggered by a seminal work by Kitaev,
who proposed an exactly solvable honeycomb model consisting of bond-directional Ising couplings,
and demonstrated that it hosts QSLs with fractionalized excitations of itinerant Majorana fermions and $\mathbb{Z}_2$ gauge fluxes \cite{Kitaev2006}.
Jackeli and Khaliullin subsequently showed that the Kitaev ($K$) interaction could be realized in alkali iridates \cite{Jackeli2009}.
However, almost all existing ``Kitaev materials" are found to exhibit long-range magnetic orderings at ambient pressure and zero magnetic field.
For example, the well-studied Na$_2$IrO$_3$ \cite{LiuBYetal2011,ChaloupkaJH2013}
and $\alpha$-RuCl$_3$ \cite{PlumbCSetal2014,KimSCKee2015,JohnsonWHetal2015} % the prominent examples
have the zigzag magnetic order at low temperatures,
while the Li$_2$IrO$_3$ family displays an incommensurate counter-rotating magnetic spiral \cite{BiffinPRL2014,WilliamsPRB2016,RousochatzakisPRB2018}.
A newly synthesized compound YbCl$_3$ with $4f$ electron configuration, which is proposed as a possible realization of the Kitaev interaction,
shows an antiferromagnetic (AFM) order with a N{\'e}el temperature $T_N = 0.60$ K \cite{XingFeng2020,SalaStone2019,HaoWo2021,SalaStone2021}.
The existence of long-range magnetic orders in these compounds is naturally understood as a consequence of non-Kitaev interactions
which contaminate the fragile Kitaev QSL.
The non-Kitaev interactions include the Heisenberg ($J$) interaction,
and also the symmetric off-diagonal $\Gamma$ and $\Gamma'$ interactions
which mainly come from the spin-orbit coupling \cite{RanLeeKeePRL2014} and trigonal distortion \cite{RauKeeArXiv2014}, respectively.

Hitherto, $\alpha$-RuCl$_3$ has drawn immense attention for the existence of fingerprints of fractionalized excitations
\cite{BanerjeeNatMat2016,BanerjeeScience2017,RanYuLiWen2017}.
Also of note is that an in-plane magnetic field of roughly 8 T can suppress the underlying magnetic order,
leading to an intermediate phase (IP) which could survive in a finite interval of magnetic field
\cite{LeahyPRL2017,SearsZhaoLynnetal2017,BaekPRL2017,Wolter2017,WangPRL2017,ZhengWenYu2017}.
However, the precise nature of this IP is still a contentious question,
with a possibility of either Majorana fermionic excitations or conventional multiparticle magnetic excitations \cite{WinterNcom2018,WulferdingNC2020}.
Noteworthily, the former scenario is in line with the tempting observation of a half-integer quantized thermal Hall effect \cite{KasaharaNature2018}.
In addition, a convictive model which harbors such an IP on top of the zigzag ordering is still absent,
although there is a consensus regarding the minimal $K$-$\Gamma$ model \cite{WangDYLi2017,SearsKim2020}.
On the other hand, when an out-of-plane magnetic field is applied, a metamagnetic transition due to the possible spin-flop process
is also reported but with a large critical magnetic field \cite{SearsKim2015}.
The fact that the discrepancy between the in-plane and out-of-plane Land{\'e} $g$-factors is modest
implies a significant role played by the symmetric off-diagonal $\Gamma$ interaction.
Meanwhile, a small $\Gamma'$ interaction stemming from the inevitable trigonal distortion should also be involved \cite{RauKeeArXiv2014}.
This term is essential for explaining the zigzag ordering in $\alpha$-RuCl$_3$ \cite{MaksimovCherny2020,AndradeJV2020},
and could enhance the mass gap of Majorana fermions generated by external magnetic fields \cite{TakikawaFuj2019,TakikawaFuj2020}.
Until now, many theoretical models such as $J$-$K$ model \cite{JiangDevJng2019},
$K$-$\Gamma$-$\Gamma'$ model \cite{GordonCSetal2019,LeeKCetal2020}, and $K$-$\Gamma$-$J_3$ model \cite{KimSota2020},
have been adopted to embrace the field-induced IPs that may relate to the experimental phenomena observed in $\alpha$-RuCl$_3$.

To study the intriguing IPs in the presence of a magnetic field,
we start from a $\Gamma$-$\Gamma'$ model with a dominant AFM $\Gamma$ interaction.
Here, the ground state is known to host two exotic phases dubbed $\Gamma$ spin liquid ($\Gamma$SL, named after the ground state of the honeycomb $\Gamma$ model \cite{LuoNPJ2021,CatunYWetal2018,GohlkeWYetal2018})
and chiral-spin ordering stabilized by a small AFM $\Gamma'$ interaction \cite{LuoStaKee2022}.
The model is equivalent to a $U(1)$-symmetric XXZ model when $\Gamma' = \Gamma$,
and the ground state turns out to be an AFM$_c$ state whose magnetic moment is along the $\textbf{c}$ [111] direction.
A natural question in mind is that if an IP could appear over the disordered phases or AFM$_c$ states in the presence of an external magnetic field.
We recall that the uniaxial Heisenberg antiferromagnet undergoes a spin-flop transition
when a magnetic field is applied parallel to the easy-axis direction \cite{AndersonCallen1964,Morrison1973}.
In the spin-flop region, the spins exhibit considerable components that are normal to the field direction,
albeit with somewhat canting toward the applied field \cite{Tian2021}.
To this end, we apply a [111] magnetic field in the $\Gamma$-$\Gamma'$ model,
and a spin-flop phase is found to set in above the $\Gamma$SL, the chiral-spin ordering, and the AFM$_c$ phase, before entering into the paramagnetic phase at large field.
Notably, the spin-flop phase in the parameter region with $\Gamma' = \Gamma$ could be interpreted as a superfluid phase
in the hard-core extended Bose-Hubbard model \cite{Wessel2007,GanWenYeetal2007}.

The rest of the paper is organized as follows.
In Sec.~\ref{SEC:MODEL}, we introduce the generic $JK\Gamma\Gamma'$ model on the honeycomb lattice,
followed by a brief mention of our numerical and theoretical methods.
In Sec.~\ref{SEC:MCvsSW}, we perform both classical and semiclassical studies of the zero-field $\Gamma$-$\Gamma'$ model,
in connection to a previous quantum study \cite{LuoStaKee2022}.
Section~\ref{SEC:QPD} presents a field-induced quantum phase diagram,
with an emphasis on the $\Gamma$SL and chiral spin state.
In Sec.~\ref{SEC:SpinFlop}, a thorough analysis of the field-induced spin-flop phase is shown.
Finally, conclusions are presented in Sec.~\ref{SEC:CONC}.

\begin{figure*}[htb]
\centering
\includegraphics[width=0.95\linewidth, clip]{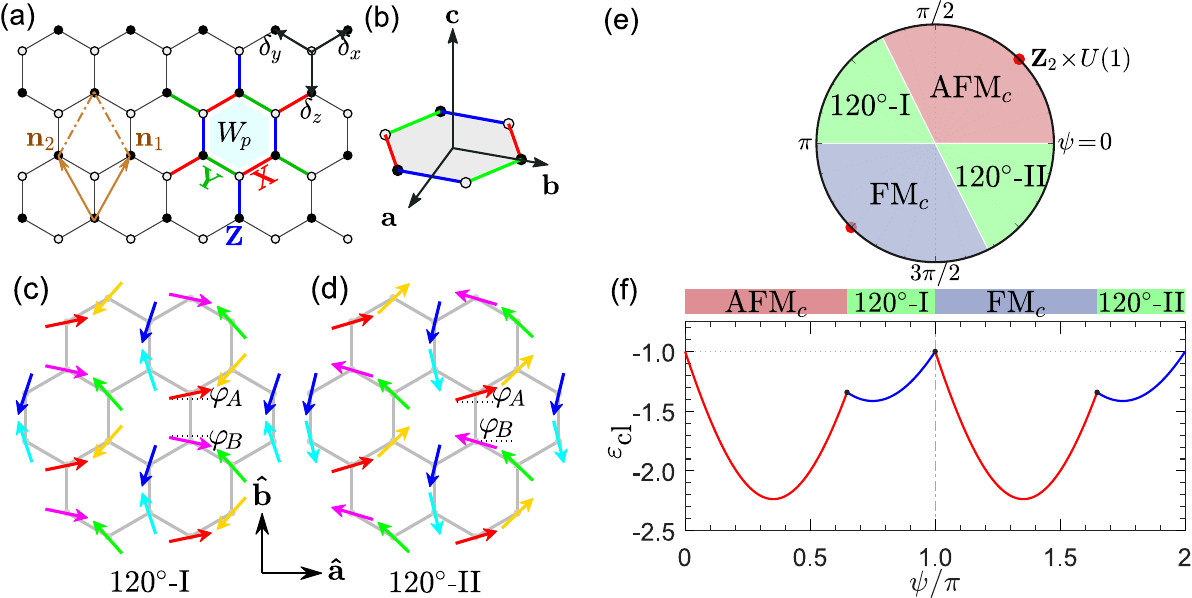}\\
\caption{(a) Honeycomb lattice with a YC structure.
    The unit cell defined by primitive vectors $\textbf{n}_{1,2} = (\pm\sqrt3/2, 3/2)$ contains two (open and filled) sites.
    $\boldsymbol{\delta}_x$ $(\sqrt3/2, 1/2)$, $\boldsymbol{\delta}_y$ $(-\sqrt3/2, 1/2)$, and $\boldsymbol{\delta}_z$ $(0, -1)$ are the unit vectors
    along the $\textbf{X}$ (red), $\textbf{Y}$ (green), $\textbf{Z}$ (blue) bonds, respectively.
    (b) Layout of the honeycomb plane spanned by ${\textbf{a}}$ [11$\bar2$] and ${\textbf{b}}$ [$\bar1$10].
    ${\textbf{c}}$ [111] represents the out-of-plane direction which is perpendicular to the honeycomb plane.
    (c) and (d) show the in-plane spin structures of the 120$^{\circ}$-I and 120$^{\circ}$-II phases, respectively.
    Here, $\varphi_A$ and $\varphi_B$ are the referring angles of $A$ and $B$ sublattices with respect to the ${\textbf{a}}$ direction.
    They satisfy the relation $\varphi_A + \varphi_B = 0$~(or $2\pi$) for 120$^{\circ}$-I phase
    and $\varphi_A + \varphi_B = \pi$~(or $3\pi$) for 120$^{\circ}$-II phase.
    (e) Classical phase diagram of the $\Gamma$-$\Gamma'$ model with $\Gamma = \cos\psi$ and $\Gamma' = \sin\psi$,
    which includes the AFM$_c$ phase, the FM$_c$ phase, and the $120^{\circ}$ phases.
    (f) The classical ground-state energy $\varepsilon_{\textrm{cl}}$ versus $\psi$.
    }\label{FIG-GemClPD}
\end{figure*}

%%%%%%%%%%%%%%%%%%%%%%%%%%%%%%%%%%%%%%%%%%%%%%%%%%%%%%%%%%%%%%%%%%%%%%%%%%%%%%
\section{Model and Methods}\label{SEC:MODEL}
%%%%%%%%%%%%%%%%%%%%%%%%%%%%%%%%%%%%%%%%%%%%%%%%%%%%%%%%%%%%%%%%%%%%%%%%%%%%%%

In the multitudinous Kitaev materials with spin-orbit coupled pseudospin-$1/2$ degrees of freedom,
the paradigmatic model takes the general $JK\Gamma\Gamma'$ form on a honeycomb lattice \cite{RanLeeKeePRL2014,RauKeeArXiv2014},
\begin{align}\label{JKGGpHc-Ham}
\mathcal{H} =
    & \sum_{\left<ij\right>\parallel\gamma} \Big[J \mathbf{S}_i \cdot \mathbf{S}_j + K S_i^{\gamma} S_j^{\gamma}
    + \Gamma \big(S_i^{\alpha}S_j^{\beta}+S_i^{\beta}S_j^{\alpha}\big)\Big]   \nonumber \\
    & + \Gamma' \sum_{\left<ij\right>\parallel\gamma}
        \Big[\big(S_i^{\alpha} + S_i^{\beta}\big) S_j^{\gamma} + S_i^{\gamma} \big(S_j^{\alpha} + S_j^{\beta}\big) \Big]    \nonumber \\
    & - \sum_i \widehat{\mathbf{h}} \cdot \mathbf{S}_i,
\end{align}
where $S_i^{\gamma}$~($\gamma$ = $x$, $y$, and $z$) is the $\gamma$-component of spin-1/2 operator at site $i$.
On $z$ bonds $(\alpha, \beta, \gamma) = (x, y, z)$, with cyclic permutation for $x$ and $y$ bonds (see Fig.~\ref{FIG-GemClPD}(a)).
$J$ and $K$ are the diagonal Heisenberg and Kitaev interactions, respectively,
while $\Gamma$ and $\Gamma'$ are the symmetry-allowed off-diagonal exchanges.
The last term in Eq.~\eqref{JKGGpHc-Ham} specifies a uniform external magnetic field $h = |\widehat{\mathbf{h}}|$
in the $[111]$ direction, which is perpendicular to the honeycomb lattice as illustrated in Fig.~\ref{FIG-GemClPD}(b).
On account of the possible microscopic Hamiltonian of $\alpha$-RuCl$_3$,
the model \eqref{JKGGpHc-Ham} has been studied previously with $K$, $\Gamma$ being treated as leading interactions \cite{GordonCSetal2019,LeeKCetal2020,ChernKLK2020}.
From a theoretical point of view, the AFM Kitaev model in a [111] magnetic field has been studied extensively and a QSL is found in an intermediate field
despite that its nature is still under study (see Ref.~\cite{ZhangHalBat2021} and references therein).
On the other hand, near the dominant AFM $\Gamma$ region, the $\Gamma$SL and the chiral spin phase are identified by tuning the $\Gamma'$ term \cite{LuoStaKee2022}.

In the subsequent sections, we will perform a hierarchical study of the $\Gamma$-$\Gamma'$ model in a [111] magnetic field
where the Heisenberg ($J$) interaction and the Kitaev ($K$) interaction are switched off.
The classical Luttinger-Tisza method is used to map out the zero-field phase diagram \cite{LuttingerTisza1946,Litvin1974},
while the classical Monte Carlo (MC) simulation is performed in the presence of a finite magnetic field \cite{Metropolis1953}.
The simulation are executed in a low-temperature range with dozens of replicas.
For each given temperature, we use the heat-bath algorithm to target the lowest energy with a MC step of five millions.
In addition, the thermal replicas where configurations swap between different temperatures are allowed
with a probability according to a detailed balance condition \cite{HukushimaNemoto1996}.
When considering the effect of quantum fluctuations,
we calculate the spin-wave energy, dispersion relations, and the Chern number
with the help of linear spin-wave theory~(LSWT) \cite{MakCh2016}.

Apart from the classical and semiclassical treatments,
this model is studied massively by the density-matrix renormalization group~(DMRG) method on two distinct cluster geometries
\cite{White1992,Peschel1999,StoudenmireWhite2012}.
The DMRG is initially invented as a powerful approach aiming to solve problems in one dimension,
and stands out as a competitive method for dealing with two-dimensional problems.
In the latter case, one needs to map the physical two-dimensional lattice to the one-dimensional chain properly.
This process will inevitably involve long-range correlation and entanglement \cite{StoudenmireWhite2012}.
However, these issues are not very severe if the number of sites is not too large or the width of the cylinder is not too big,
and could be reduced essentially by increasing the block states and performing finite-size scaling.
We focus primarily on a 24-site $C_6$-symmetric hexagonal cluster under full periodic boundary condition,
and the method to map it to a one-dimensional chain is shown in the Supplemental Material \cite{SuppMat}.
In addition, we also consider the $L_x \times L_y$ YC cluster under cylindrical boundary condition with total sites $N = L_xL_y$ (cf. Fig.~\ref{FIG-GemClPD}(a)).
During the calculation, the truncation error will change as we scan the superblock and it also decreases with the increase of the block state.
Therefore, we keep as many as $m$ = 3000 block states and perform up to 12 sweeps until the worst truncation error is smaller than $10^{-6}$.

\section{Classical and semiclassical study of the $\Gamma$-$\Gamma'$ model}\label{SEC:MCvsSW}

\subsection{Lutinger-Tisza analysis}

Before presenting the quantum study of the $\Gamma$-$\Gamma'$ model, it is helpful to have a look at the classical phase diagram.
The Luttinger-Tisza method has been demonstrated to be powerful for the determination of magnetic ground states in various classical spin models \cite{LuttingerTisza1946,Litvin1974}.
In these models, the classical spins are treated as $O(3)$ vectors which satisfy the condition $|\mathbf{S}_i|^2 = S^2$.
In the spirit of the Luttinger-Tisza method, this `hard constraint' is replaced by a `soft constraint' $\sum_i |\mathbf{S}_i|^2 = NS^2$ tentatively,
and the authentic ground state is selected from those solutions derived under the soft constraint that additionally meets the hard constraint.
Successful applications of the Luttinger-Tisza method to the spin-orbit coupled model \eqref{JKGGpHc-Ham} in some special cases are shown previously \cite{ChalKhal2015,Rousochatzakis2017}.

We choose the primitive vectors of the honeycomb lattice as $\textbf{n}_{1,2} = (\pm\sqrt3/2, 3/2)$
(see Fig.~\ref{FIG-GemClPD}(a)),
and the sites are represented as ($\bm{R}$, $\upsilon$),
where $\bm{R}$ marks the position of the unit cell and $\upsilon$~(= 1, 2) is the sublattice index.
Transforming the spin operators via
$\vec{S}_{\bm{R},\upsilon}\!=\!\sum_{\bm{q}} e^{i \bm{q}\cdot\bm{R}} \vec{S}_{\bm{q},\upsilon}$ with $\vec{S}_{\bm{q},\upsilon}\!=\!(S_{\bm{q},\upsilon}^x,S_{\bm{q},\upsilon}^y,S_{\bm{q},\upsilon}^z)^T$,
we cast the entire Hamiltonian in the reciprocal space as
\begin{equation}\label{eq:Hq}
\mathcal{H}/N=\frac{1}{4}\sum_{\bm{q}} \left(\vec{S}_{-\bm{q},1}^T,\vec{S}_{-\bm{q},2}^T\right) \cdot
\boldsymbol{\Lambda}_{\bm{q}} \cdot
\left(\begin{array}{c}
\vec{S}_{\bm{q},1} \\
\vec{S}_{\bm{q},2}
\end{array}\right),
\end{equation}
where the 6$\times$6 interaction matrix $\boldsymbol{\Lambda}_{\bm{q}}$ is an anti-diagonal block matrix
\begin{equation*}
\boldsymbol{\Lambda}_{\bm{q}} = \left(
\begin{array}{cc}
\vec{0}                             & \boldsymbol{\Delta}_{\bm{q}} \\
\boldsymbol{\Delta}^\ast_{\bm{q}}   & \vec{0}
\end{array}
\right).
\end{equation*}
Here,
\begin{equation*}
\boldsymbol{\Delta}_{\bm{q}} \!= \left(\!\!
\begin{array}{ccc}
\mathcal{K}(\varsigma_1)        &       \mathcal{T}(1)                  &       \mathcal{T}(\varsigma_2)        \\
\mathcal{T}(1)                  &       \mathcal{K}(\varsigma_2)        &       \mathcal{T}(\varsigma_1)        \\
\mathcal{T}(\varsigma_2)        &       \mathcal{T}(\varsigma_1)        &       \mathcal{K}(1)        \\
\end{array}\!\!
\right),
\end{equation*}
\normalsize
with
$\mathcal{K}(\varsigma) = J\gamma_{\bm{k}} + \varsigma K$
and
$\mathcal{T}(\varsigma) = \Gamma'\gamma_{\bm{k}} + \varsigma(\Gamma-\Gamma')$.
The momentum-dependent arguments read as
\begin{equation*} % \label{VarSigma12}
\varsigma_{1,2} = e^{-\ii\bm{q}\textbf{n}_{1,2}} = e^{-\ii(\pm\sqrt{3}q_x+3q_y)/2}
\end{equation*}
and
\begin{equation*} % \label{EtaK}
\gamma_{\bm{q}} = 1+ \varsigma_{1} + \varsigma_{2} = 1 + 2\cos\frac{\sqrt{3}q_x}{2} e^{-\ii 3q_y/2}.
\end{equation*}

According to the Luttinger-Tisza minimization,
the lowest eigenvalue of $\boldsymbol{\Lambda}_{\bm{q}}$ in the entire Brillouin zone provides a lower bound of the classical energy.
Noticing that
$\mathcal{D}_{\bm{q}} = \boldsymbol{\Delta}^{*}_{\bm{q}}\boldsymbol{\Delta}_{\bm{q}} = \boldsymbol{\Delta}^{\dagger}_{\bm{q}}\boldsymbol{\Delta}_{\bm{q}}$,
we find that
\begin{equation}\label{EgLT}
E/(N S^2) \ge -\frac{\sqrt{\lambda_{\max}}}{2}
\end{equation}
where $\lambda_{\max}$ is the maximal eigenvalue of $\mathcal{D}_{\bm{q}}$ at the corresponding ordering wave vector $\mathbf{Q}$.
The magnetic moment direction $\vec{m}$ can be obtained afterwards by checking the spin-length constraint.

We have applied the Luttinger-Tisza method to the $\Gamma$-$\Gamma'$ model,
and the corresponding classical phase diagram is found to include an AFM$_c$ phase when $\Gamma' > 0$ and a ferromagnetic (FM$_c$) phase when $\Gamma' < 0$.
Here, the subscript $c$ represents that magnetic moment direction is along the ${\textbf{c}}$ [111] direction.
The energy of the AFM$_c$ phase is $\varepsilon_{\rm cl} = -(\Gamma+2\Gamma')$, while it is $\Gamma+2\Gamma'$ for FM$_c$ phase.
The classical phase diagram also contains two $120^{\circ}$ phases but with different relative angles (see Fig.~\ref{FIG-GemClPD}(c) and (d)).
For the 120$^{\circ}$ phases, all spins lie in the $ab$-plane and are divided into two interpenetrating
parts on $A$ and $B$ sublattices of honeycomb lattice,
where on each sublattice the spins on the corner of an equilateral triangle are mutually oriented to each other with 120 angles.
Assuming that $\varphi_A$ and $\varphi_B$ are the in-plane angles of $A$ and $B$ sublattices with respect to the ${\textbf{a}}$ direction,
then the classical energy per site is given by
\begin{equation}\label{EQ:ClEg120}
\varepsilon_{\textrm{cl}}^{120^{\circ}} = -(\Gamma-\Gamma')\cos\big(\varphi_A + \varphi_B - \pi\big).
\end{equation}
The optimal angles of $\varphi_A$ and $\varphi_B$ depend on the sign of $\Gamma-\Gamma'$,
where a negative sign denotes $\varphi_A+\varphi_B = 0$ or $2\pi$ (see Fig.~\ref{FIG-GemClPD}(c)),
while a positive sign represents $\varphi_A+\varphi_B = \pi$ or $3\pi$ (see Fig.~\ref{FIG-GemClPD}(d)).
There is no extra restriction on the values of $\varphi_A$ and $\varphi_B$,
implying an emergent $U(1)$ symmetry in the $ab$ plane.
In both cases, we have the classical energy as
$\varepsilon_{\textrm{cl}}^{120^{\circ}} = -\left|\Gamma-\Gamma'\right|$.

In addition, we also parameterize $\Gamma = \cos\psi$ and $\Gamma' = \sin\psi$,
and show the classical phase diagram in Fig.~\ref{FIG-GemClPD}(e).
It can be found that the AFM$_c$ phase is preferred when $\psi \in (0, \psi_0)$ with $\psi_0 = \pi - \tan^{-1}(2) \approx 0.6476\pi$
and occupies nearly one third of the whole $\psi$ circle.
Remarkably, when $\psi = \pi/4$ (i.e., $\Gamma = \Gamma'$), the $\Gamma$-$\Gamma'$ model can be reduced to
\begin{align}\label{EQ:XXZG-Ham}
\mathcal{H} =& \Gamma \sum_{\left<ij\right>} \Big[-\frac12(\tilde{S}_i^+\tilde{S}_j^- + \tilde{S}_i^-\tilde{S}_j^+) + 2\tilde{S}_i^z\tilde{S}_j^z\Big],
\end{align}
which is nothing but an \textit{easy-axis} XXZ model with a $\mathbb{Z}_2 \ltimes U(1)$ symmetry.
The AFM$_c$ phase and the FM$_c$ phase are smoothly connected to the hidden $SU(2)$ Heisenberg model
at $(J, \Gamma, \Gamma') = (1/3,1/3,1/3)$ and $(J, \Gamma, \Gamma') = (1/3,-2/3,-2/3)$ \cite{ChalKhal2015}, respectively.
Besides, from the pinnacles of the energy curve shown in Fig.~\ref{FIG-GemClPD}(f) we can tell that all the classical phase transitions are of first order.
In what follows, we take $\Gamma = 1$ as the energy unit.

\begin{figure}[!ht]
\centering
\includegraphics[width=0.95\columnwidth, clip]{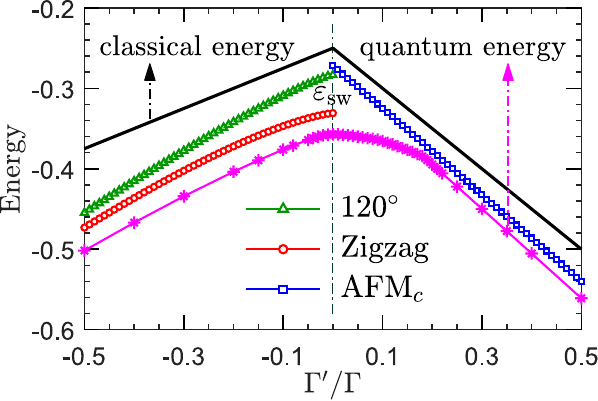}\\
\caption{The spin-wave energy $\varepsilon_{\rm sw}$ for the zigzag phase (red circle), 120$^{\circ}$ phase (green triangle), and AFM$_c$ phase (blue square)
    in the $\Gamma$-$\Gamma'$ model. The classical (black line) and quantum (pink diamond) energy per site with $S = 1/2$ are also shown for comparison.}\label{FIG-SWEg}
\end{figure}

\subsection{Linear spin-wave theory}

Classically, there is a direct 120$^{\circ}$-AFM$_c$ transition in the vicinity of the AFM $\Gamma$ limit as $\Gamma'$ is varied.
The quantum fluctuation manifests its effect by altering the underlying phases at least in two aspects \cite{LuoStaKee2022}.
One is that the 120$^{\circ}$ phase is replaced by the zigzag phase when $\Gamma'$ is negative.
The other is that, for small but positive $\Gamma'$ interaction, there are two exotic phases which are intervened between the magnetically ordered states.
Here we show that the LSWT is amenable to illuminate the effect of quantum fluctuation.
Within the framework of LSWT, the quadratic Hamiltonian in the momentum space reads \cite{MakCh2016}
\begin{eqnarray}\label{EQ:H120SW}
\overline{\mathcal{H}}
&=& NS(S+1) \varepsilon_{\rm cl} + \frac{S}{2} \sum_{\bm q}
    \hat{\bf x}_{\bm q}^\dagger
    \hat{\bf H}_{\bm q}
    \hat{\bf x}_{\bm q}^{\phantom{\dagger}},
\end{eqnarray}
where $\hat{\bf x}^\dag_{\bm q}=\big( a^\dag_{1, \bm q}, a^\dag_{2, \bm q}, \cdots, a_{1, -\bm q},b_{2, -\bm q}, \cdots\big)$ is the Nambu spinor
and $\hat{\bf H}_{\bm q}$ is a $2\times 2$ block matrix termed Bogliubov-de Gennes (BdG) Hamiltonian.
The length of Nambu spinor should be $2n_s$ where $n_s$ is the number of sites in one unit cell.
The bosonic BdG Hamiltonian is diagonalized via a paraunitary matrix $\bm{T}_{\bm{q}}$,
\begin{eqnarray}\label{EQ:BdGDiag}
\bm{T}_{\bm{q}}^{\dagger} \hat{\bf H}_{\bm{q}} \bm{T}_{\bm{q}} =
    \left(\begin{array}{@{}cc@{}}
	\boldsymbol{\Omega}_{\bm{q}}   &   0 \\
	0                              &   \boldsymbol{\Omega}_{-\bm{q}}
    \end{array}\right),
\end{eqnarray}
where $\boldsymbol{\Omega}_{\bm{q}} = \textrm{diag}\big(\omega_{\bm{q}1}, \omega_{\bm{q}2}, \cdots, \omega_{\bm{q}n_s}\big)$
whose diagonal elements are the magnon dispersions $\omega_{\bm{q}\upsilon}$ ($\upsilon = 1, 2, \cdots, n_s$).
The paraunitary matrix $\bm{T}_{\bm{q}}$ satisfies the boson relation
\begin{eqnarray}\label{EQ:TranTSigma}
\bm{T}_{\bm{q}}^{\dagger} \boldsymbol{\Sigma} \bm{T}_{\bm{q}} =
\bm{T}_{\bm{q}} \boldsymbol{\Sigma} \bm{T}_{\bm{q}}^{\dagger} = \boldsymbol{\Sigma},
\end{eqnarray}
where $\boldsymbol{\Sigma} = \mathrm{diag}(+\hat{\boldsymbol{1}}, -\hat{\boldsymbol{1}})$.
In other words, the magnon dispersions $\omega_{\bm{q}\upsilon}$ can also be determined by diagonalizing $\boldsymbol{\Sigma} \hat{\bf H}_{\bm q}$.
The spin-wave energy is then given by
\begin{eqnarray}\label{EQ:SpinWaveEg}
\varepsilon_{\rm sw} = S(S+1) \varepsilon_{\textrm{cl}} + \frac{S}{2n_s} \sum_{\{\upsilon\}\in n_s} \int \frac{d^2\mathbf q}{(2\pi)^2} \omega_{\bm{q}\upsilon}.
\end{eqnarray}

Figure~\ref{FIG-SWEg} shows the spin-wave energy $\varepsilon_{\rm sw}$ for the zigzag phase, 120$^{\circ}$ phase,
and AFM$_c$ phase in the window of $-0.5 \leq \Gamma'/\Gamma \leq 0.5$.
When $\Gamma' < 0$, energy of the zigzag phase is considerably smaller than that of the 120$^{\circ}$ phase,
showing that the quantum fluctuation would provoke the zigzag ordering as the true ground state.
In the neighboring of the AFM $\Gamma$ limit, magnon gap $\Delta$ of the zigzag phase decreases gradually and vanishes when $\Gamma'/\Gamma = 0$ (not shown).
This phenomenon is called the magnon instability and is a signature of phase transition \cite{Maksimov2019}.
Hence, the zigzag phase can not surpass the line of $\Gamma'/\Gamma = 0$ and thus cannot survive in the presence of an AFM $\Gamma'$ term.
Whereas the AFM$_c$ phase is favored for modest positive $\Gamma'$ interaction,
there is a noteworthy energy jump between the zigzag phase and the AFM$_c$ phase near $\Gamma'/\Gamma = 0$.
Our spin-wave result implies that an intermediate region should exist as a consequence of competing interactions.
The classical (black line) and quantum (pink diamond) energy per-site are also shown in Fig.~\ref{FIG-SWEg} for comparison.
It is observed that the spin-wave energy is lower than the classical energy, but is higher than the quantum case.

%%%%%%%%%%%%%%%%%%%%%%%%%%%%%%%%%%%%%%%%%%%%%%%%%%%%%%%%%%%%%%%%%%%%%%%%%%%%%%%%%%%%%%%%%%%%%%%%%%%%%
\section{Magnetic field-induced quantum phase diagram}\label{SEC:QPD}
%%%%%%%%%%%%%%%%%%%%%%%%%%%%%%%%%%%%%%%%%%%%%%%%%%%%%%%%%%%%%%%%%%%%%%%%%%%%%%%%%%%%%%%%%%%%%%%%%%%%%

\begin{figure}[!ht]
\centering
\includegraphics[width=0.95\columnwidth, clip]{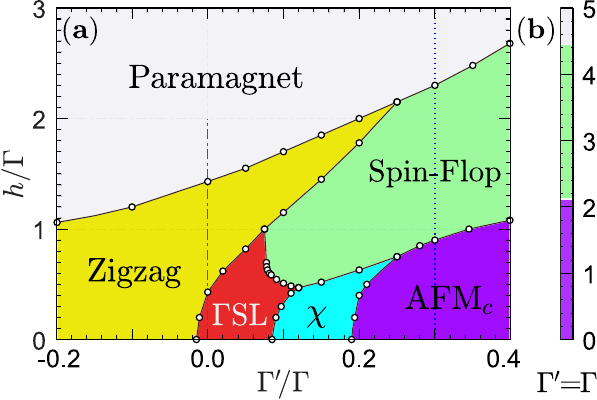}\\
\caption{(a) Quantum phase diagram of the $\Gamma$-$\Gamma'$ model in a [111] magnetic field.
    The DMRG calculation is performed on a 24-site hexagonal cluster.
    The phase diagram contains a zigzag phase, a $\Gamma$SL, a chiral spin state (marked by the symbol $\chi$), and an AFM$_c$ phase at zero field.
    A spin-flop phase and a paramagnetic phase are induced upon increasing the magnetic field.
    We note that the precise phase boundary between the zigzag phase and the paramagnetic phase is smeared
    due to the unavoidable multi-step magnetization before entering into the totally polarized region.
    The vertical cuts along the lines of $\Gamma'/\Gamma = 0.0$ (red dash-dotted line) and $\Gamma'/\Gamma = 0.3$ (blue dotted line)
    are the main focuses of the following study.
    The study of other cuts are shown in the Supplemental Material \cite{SuppMat}.
    (b) Quantum phase transitions along the line of $\Gamma'/\Gamma = 1$.
    The low-field transition at $h_{t1} = 2.115(3)$ is first order, while the high-field transition at $h_{t2} \approx 4.50$ seems to be continuous.
    }\label{FIG-FldhcPD}
\end{figure}

In a previous study of the $\Gamma$-$\Gamma'$ model by the authors \cite{LuoStaKee2022},
it is shown that there is indeed an intermediate region between the zigzag phase and the AFM$_c$ phase at the quantum level.
In the range of $-0.015 \lesssim \Gamma'/\Gamma \lesssim 0.095$, there is a gapless $\Gamma$SL which is characterized by a hidden plaquette correlation \cite{LuoNPJ2021}.
Besides, a chiral-spin ordered state with spontaneously time-reversal symmetry breaking appears when $0.095 \lesssim \Gamma'/\Gamma \lesssim 0.185$.
Here, we go beyond that study by applying an out-of-plane magnetic field,
and the resulting phase diagram is shown in Fig.~\ref{FIG-FldhcPD}.
The DMRG computation is mainly executed in the 24-site hexagonal cluster.
We have also checked the phase diagram on the \textrm{YC6} cylinder of $12\times6$,
which basically remains unchanged despite a tiny shift of the phase boundaries.

Throughout the phase diagram, there are six distinct phases and two of them only exist in the presence of a finite magnetic field.
One is a conventional paramagnetic phase,
while the other is a spin-flop phase which also exhibits an in-plane magnetization when compared with the paramagnetic phase.
Starting from the magnetically ordered states at zero field,
transition between the zigzag phase and the paramagnetic phase is first order, as reflected by the jump in the magnetic order parameter.
By contrast, the spin-flop phase is sandwiched between the AFM$_c$ phase and the paramagnetic phase.
We note that the spin-flop phase has an intimate relation to the superfluid phase identified in the extended Bose-Hubbard model \cite{Wessel2007,GanWenYeetal2007}.
In addition, the regions of $\Gamma$SL and chiral spin state are enlarged but are terminated before entering into the paramagnetic phase.
In what follows we will concentrate on the $\Gamma$SL and chiral spin state,
while leaving the discussion on the spin-flop phase to the next section.

\subsection{$\Gamma$SL in the magnetic field}

\begin{figure}[tb]
\centering
\includegraphics[width=0.95\columnwidth, clip]{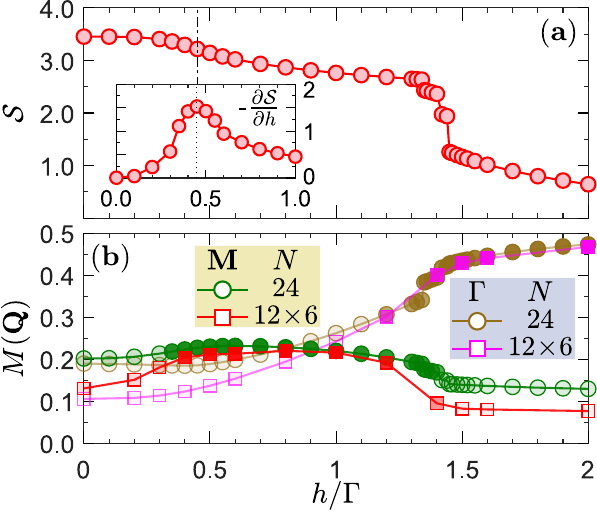}\\
\includegraphics[width=0.95\columnwidth, clip]{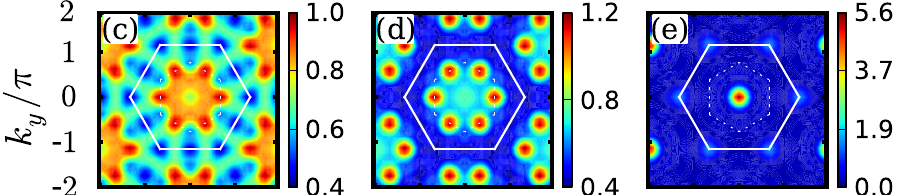}\\
\includegraphics[width=0.95\columnwidth, clip]{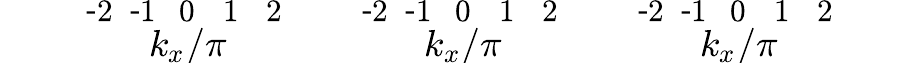}\\
\caption{(a) Behavior of the entanglement entropy $\mathcal{S}$ as a function of magnetic field $h$ in the $\Gamma$-$\Gamma'$ model with $\Gamma'/\Gamma = 0$.
    The underlying geometry is a 24-site hexagonal cluster. Inset: The first-order derivative of $\mathcal{S}$ with respect to $h$.
    (b)Magnetic order parameters $M_N(\textbf{Q})$ for the zigzag order and paramagnetic phase
    with $\textbf{Q} = \textrm{\bf{M}}$ and $\boldsymbol{\Gamma}$, respectively.
    The geometries are a 24-site hexagonal cluster and a $12\times6$ YC cylinder.
    Panels (c-e) show the static structure factor at a field of (c) $h = 0.2$ ($\Gamma$SL), (d) $h = 1.0$ (zigzag phase),
    and (e) $h = 2.0$ (paramagnetic phase), respectively.
    }\label{FIG-HCGAFld}
\end{figure}

We start from the $\Gamma$SL and investigate its fate in the presence of a magnetic field.
The von Neumann entanglement entropy is a reliable quantity to capture the phase transitions between the phases with unique ground states.
It is defined as $\mathcal{S} = -\mathrm{tr}(\rho_s\ln\rho_s)$ where $\rho_s$ is the reduced density matrix of one half of the system \cite{EisertRMP2010}.
$\mathcal{S}$ displays a jump at the transition point if the transition is first order,
otherwise it varies smoothly with the driving parameter.
Figure~\ref{FIG-HCGAFld}(a) shows the behavior of entanglement entropy $\mathcal{S}$ on the 24-site cluster.
When we apply a small magnetic field, entanglement entropy is maintained around 3.5,
followed by a sustaining decrease with a steepest drop at $h_{t1} \approx 0.45(5)$ (see inset).
The entanglement entropy does not experience a big change until an abrupt reduction around $h_{t2} \approx 1.40(5)$.
The consecutive release of entropy therein may imply a multi-step alignment of the spins towards a more parallel structure in the paramagnetic phase.
We expect the interval of this metastate shrinks with the increase of the system size.

To figure out the nature of the intermediate region, we resort to the static structure factor~(SSF)
$\mathbb{S}_{N,\tau}({\bm{q}}) = \sum_{\alpha\beta}\delta_{\alpha\beta}\mathbb{S}_{N,\tau}^{\alpha\beta}({\bm{q}})$
where
\begin{equation}\label{EQ:SSF}
\mathbb{S}_{N,\tau}^{\alpha\beta} = \frac{1}{N}\sum_{ij}
\big(\langle S^{\alpha}_i S^{\beta}_j\rangle  - \tau \langle S^{\alpha}_i\rangle \langle S^{\beta}_j\rangle \big)
e^{i{\bm{q}}\cdot{({\bm{R}}_i-{\bm{R}}_j)}}.
\end{equation}
Here, ${\bm{R}}_i$ is the position of site $i$ and $\bm{q}$ is the wavevector in the reciprocal space.
The symbol $\tau$ could be either 0 or 1, and it indicates that the effect of magnetic field is either kept or deducted, respectively.
We note that when $\tau = 1$, only the intensity around the very center (i.e., $\boldsymbol{\Gamma}$ point) in the Brillouin zone is reduced.
Figure~\ref{FIG-HCGAFld}(c-e) show the snapshots of SSF $\mathbb{S}_{N,\tau}({\bf{Q}})$
in a field of $h = 0.1$ ($\tau = 1$), 1.0 ($\tau = 1$), and 2.0 ($\tau = 0$), respectively.
In Fig.~\ref{FIG-HCGAFld}(c), intensity in the reciprocal space is very diffusive,
albeit with a soft peak at $\bf{M}$ point that may relate to the adjacent zigzag ordering.
By contrast, a sharp peak at $\bf{M}$ point could be spotted in the intermediate region as shown in Fig.~\ref{FIG-HCGAFld}(d).
Upon applying a higher magnetic field, there is a paramagnetic phase which displays a visible peak at $\boldsymbol{\Gamma}$ point (see Fig.~\ref{FIG-HCGAFld}(e)).
We define the order parameter $M_{N,\tau}({\bf{Q}}) = \sqrt{\mathbb{S}_{N,\tau = 0}({\bf{Q}})/N}$
with ${\bf{Q}}$ being the ordering wavevector.
In Fig.~\ref{FIG-HCGAFld}(b) we show the order parameters $M_N(\textbf{Q})$
of the zigzag phase ($\textbf{Q} = \textbf{M}$) and paramagnetic phase ($\textbf{Q} = \boldsymbol{\Gamma}$)
on a 24-site hexagonal cluster and a $12\times6$ YC cylinder.
When $h < h_{t,1}$, the order parameter $M_N(\textbf{M})$ exhibits a considerable reduction with the increase of the system size.
Although we do not make an extrapolation of this order parameter for the lack of large clusters,
the magnetic order is likely to vanish as $N$ increases, and the low-field region should be identical to the $\Gamma$SL
identified in the zero-field study \cite{LuoNPJ2021}.
On the other hand, the ground state at $h > h_{t,2}$ is a paramagnetic phase with a almost saturated magnetic moment.
However, the most inspiring observation is that there is a zigzag ordering,
which is smoothly connected to the zigzag phase induced by the FM $\Gamma'$ interaction,
in the intermediate region of $h_{t,1} < h < h_{t,2}$.
The zigzag phase is unusual in that it only has a unique ground state with a small excitation gap.

\begin{figure}[tb]
\centering
\includegraphics[width=0.95\columnwidth, clip]{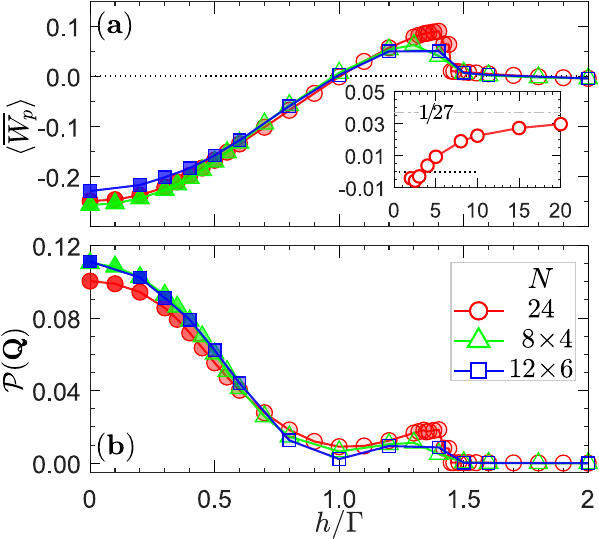}\\
\caption{(a) Flux-like density $\langle\overline{W}_p\rangle$ and (b) plaquette order parameter $\mathcal{P}(\textbf{Q})$
    on the 24-site hexagonal cluster (red circle) and the YC cylinders of $8\times4$ (green triangle) and $12\times6$ (blue square).
    The inset of (a) shows $\langle\overline{W}_p\rangle$ at large field, which tends to approach $1/27$ as field increases.
    }\label{FIG-HCGAFluxWp}
\end{figure}

\begin{figure*}[htb]
\centering
\includegraphics[width=0.95\linewidth, clip]{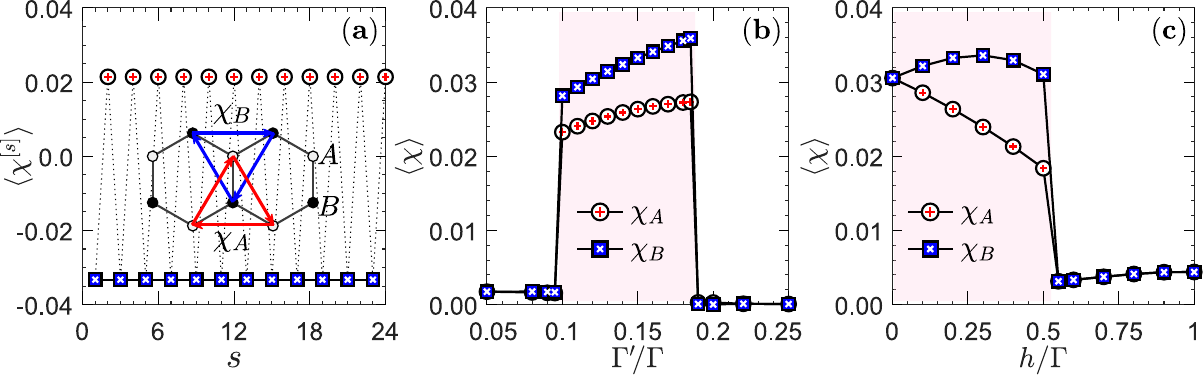}\\
\caption{(a) Chiral distribution of $\langle\chi_{ijk}^{[s]}\rangle$ as a function of site index $s$ on a 24-site hexagonal cluster with $\Gamma'/\Gamma$ = 0.15 and $h = 0.2$.
    Here, $s$ is the center site of the equilateral triangle formed by $(i,j,k)$.
    When $s$ is 2, 4, 6, $\cdots$ (red circle), $(i,j,k)$ should be odd and belongs to the $A$ sublattice
    with the chiral order parameter $\langle\chi_A\rangle = \vert\langle\chi^{[s]}\rangle\vert$ ($s$ is even).
    Otherwise, when $s$ is 1, 3, 5, $\cdots$ (blue square), $(i,j,k)$ belongs to the $B$ sublattice
    with the chiral order parameter $\langle\chi_B\rangle = \vert\langle\chi^{[s]}\rangle\vert$ ($s$ is odd).
    (b) Behaviors of $\langle\chi_A\rangle$ (red circle) and $\langle\chi_B\rangle$ (blue square) as functions of $\Gamma'/\Gamma$ with fixed $h = 0.2$.
    (c) Behaviors of $\langle\chi_A\rangle$ (red circle) and $\langle\chi_B\rangle$ (blue square) as functions of $h$ with fixed $\Gamma'/\Gamma = 0.15$.
    }\label{FIG-ChiAB}
\end{figure*}

Similar to the Kitaev honeycomb model, in the $\Gamma$-$\Gamma'$ model we also calculate the hexagonal plaquette operator \cite{Kitaev2006}
\begin{equation}\label{EQ:FluxWp}
\hat{W}_p = 2^6 S_1^{x}S_2^{y}S_3^{z}S_4^{x}S_5^{y}S_6^{z},
\end{equation}
where the sites 1--6 form a hexagon plaquette labeled by $p$ (see Fig.~\ref{FIG-GemClPD}(a)).
Without loss of generality, we define the flux-like density
$\overline{W}_p = \sum_p\langle\hat{W}_p\rangle/N_p$ where $N_p = N/2$ is the number of plaquette.
Figure~\ref{FIG-HCGAFluxWp}(a) shows the flux-like density $\overline{W}_p$ with respect to the magnetic field.
In the $\Gamma$SL region the net flux $\overline{W}_p$ is $-0.25(2)$ at zero field,
followed by a steady ascent as the field increases.
The flux-like density finally becomes positive and reach its maximal value of 0.10(1) at $h_{t,2}$.
After that one enters into the paramagnetic phase accompanied by a sudden drop of $\overline{W}_p$.
It is interesting to note that the flux-like density in the paramagnetic phase does not has a monotonous behavior;
instead, it first declines with the field and then increases again, reaching a saturated value ultimately.
For large enough magnetic field along the [111] direction,
all the spins $\mathbf{S}_i = \boldsymbol{\sigma}_i/2$ are totally polarized with the same magnitude,
$\langle \sigma_i^x\rangle$ = $\langle \sigma_i^y\rangle$ = $\langle \sigma_i^z\rangle$ = ${1}/{\sqrt3}$.
Thus, expectation value of hexagonal plaquette operator at large enough magnetic field is
\begin{equation}\label{EQ:WpPlrzd}
\langle \hat{W}_p\rangle = \left(\frac{1}{\sqrt3}\right)^6 = \frac{1}{27} \approx 0.037037\cdots.
\end{equation}
As can be seen from the inset of Fig.~\ref{FIG-HCGAFluxWp}(a),
$\overline{W}_p$ indeed approaches to $1/27$ with the increase of magnetic field.

We continue the discussion of hexagonal plaquette operator by calculating the plaquette-plaquette correlation $\langle W_p\cdot W_q\rangle$.
The plaquette structure factor is defined as \cite{SahaFZetal2019}
\begin{equation}\label{EQ:VisonSF}
\mathcal{W}_{N_p}({\bm{q}})=\frac{1}{N_p}\sum_{pq}
\langle{\hat{W}_p \hat{W}_q}\rangle e^{i{\bm{q}}\cdot{({\bm{R}}_p-{\bm{R}}_q)}},
\end{equation}
where $\bm{R}_p$ is the central position of each plaquette which forms a triangular lattice with a lattice constant of $\sqrt3$.
In the totally polarized phase, $\langle\hat{W}_p\cdot\hat{W}_q\rangle$ can only take three different values, depending on their relative positions.
If $\hat{W}_p$ and $\hat{W}_q$ are identical or totally irrelevant without any shared edge,
then $\langle\hat{W}_p\cdot\hat{W}_q\rangle$ is 1 and $1/3^6$, respectively.
Otherwise, $\hat{W}_p$ and $\hat{W}_q$ have a sole shared edge and $\langle\hat{W}_p\cdot\hat{W}_q\rangle$ = $1/3^5$.
Taken together, we have
\begin{eqnarray}\label{EQ:PlqttSFRelation}
\mathcal{W}_{N_p}(\boldsymbol{\Gamma}) = \Big(1 + \frac{11}{3^6}\Big) + \frac{N_p}{3^6}.
\end{eqnarray}
Typically, the first term in the right-hand side is dominant when $N \lesssim 100$.
To reduce the strong finite-size effect, we introduce the following plaquette order parameter
\begin{equation}\label{EQ-HiddenOPPlqtt}
\mathcal{P}_{N_p} = \sqrt{\frac{\mathcal{W}_{N_p}({\bf{Q}})}{N_p}} - \frac{1}{\sqrt{N_p}}.
\end{equation}
Figure~\ref{FIG-HCGAFluxWp}(b) shows the plaquette order parameter $\mathcal{P}(\textbf{Q})$
with the high-symmetry point $\textbf{Q} = \boldsymbol{\Gamma}$ being the center of the Brillouin zone.
In the $\Gamma$SL and the zigzag phase, $\mathcal{P}(\boldsymbol{\Gamma})$ is nonzero as the spins are noncollinear.
Furthermore, $\mathcal{P}(\boldsymbol{\Gamma})$ is more pronounced in the $\Gamma$SL,
highlighting the unusual spin pattern due to the intrinsic frustration.

\subsection{Chiral spin state}

As pointed out in Ref.~[\onlinecite{LuoStaKee2022}],
the chiral-spin ordered state could be stabilized by a small AFM $\Gamma'$ interaction that is one order of magnitude smaller than the dominated $\Gamma$ interaction.
It is known to break time-reversal symmetry spontaneously and thus exhibits a finite scalar spin chirality defined as
\begin{equation}\label{EQ:ChiIJK}
\hat{\chi}^{\triangle}_{ijk} = \hat{\mathbf{S}}_i\cdot(\hat{\mathbf{S}}_j\times\hat{\mathbf{S}}_k),
\end{equation}
where ($i, j, k$) label lattice sites of either $A$ or $B$ sublattice, forming an equilateral triangle $\triangle$ in the clockwise direction,
see inset of Fig.~\ref{FIG-ChiAB}(a).
We find that the chiral spin state could still survive up to a small magnetic field, before entering into a partially polarized phase.
Following the analysis in Ref.~[\onlinecite{LuoStaKee2022}], we focus on a point at ($\Gamma'$ = 0.15, $h$ = 0.2) in the $\Gamma$-$\Gamma'$ model,
and the distribution of chirality $\hat{\chi}$ within the 24-site cluster is shown in Fig.~\ref{FIG-ChiAB}(a).
It is clear seen that the scalar spin chirality is uniformly distributed in each sublattice and possesses an opposing sign in the $A$ and $B$ sublattices.
In addition, magnitudes of the chirality in the $A$ and $B$ sublattices,
whose absolute values are $\langle{\chi}_A\rangle \approx 0.0264$ and $\langle{\chi}_{B}\rangle \approx 0.0333$,
are no longer the same due to the existence of the magnetic field.

Figure~\ref{FIG-ChiAB}(b) presents the chiral order parameters $\langle{\chi}_A\rangle$ and $\langle{\chi}_B\rangle$ as functions of $\Gamma'$.
The chiral order parameters are very robust in the window of $0.095 \lesssim \Gamma'/\Gamma \lesssim 0.185$ and undergo drastic jumps on the brink of phase boundaries.
We also show the evolutions of chiral order parameters with respect to the magnetic field in Fig.~\ref{FIG-ChiAB}(c).
It is found that $\langle{\chi}_B\rangle$ is slightly elevated with the increase of magnetic field and displays a maximum when $h \approx 0.3$.
By contrast, $\langle{\chi}_A\rangle$ decreases almost linearly from 0.0333 (at $h = 0.0$) to 0.0195 (at $h_t \approx 0.53$).
Again, the chiral order parameters experience jumps to a small but finite value at $h > h_t$,
and the system enters into the spin-flop phase where the discrepancy between $\langle{\chi}_A\rangle$ and $\langle{\chi}_B\rangle$ disappears.

According to our previous work, the chiral spin state is known as a magnetically disordered state without long-range magnetic ordering \cite{LuoStaKee2022}.
In that study, we proposed that it could be either a gapless chiral spin liquid because of the continuous feature of the dynamic structure factor in the low frequency region,
or a symmetry-protected phase with short-range entanglement based on the modular matrix.
However, a decisive conclusion could hardly be made due to the low symmetry of the Hamiltonian and the capacity of the numerical calculation.
Hence, determining the nature of the chiral spin state is a tempting open question to be explored.

%%%%%%%%%%%%%%%%%%%%%%%%%%%%%%%%%%%%%%%%%%%%%%%%%%%%%%%%%%%%%%%%%%%%%%%%%%%%%%
\section{Field-induced spin-flop phase}\label{SEC:SpinFlop}
%%%%%%%%%%%%%%%%%%%%%%%%%%%%%%%%%%%%%%%%%%%%%%%%%%%%%%%%%%%%%%%%%%%%%%%%%%%%%%

\subsection{Overview of the classical analysis}

\begin{figure}[!ht]
\centering
\includegraphics[width=0.95\columnwidth, clip]{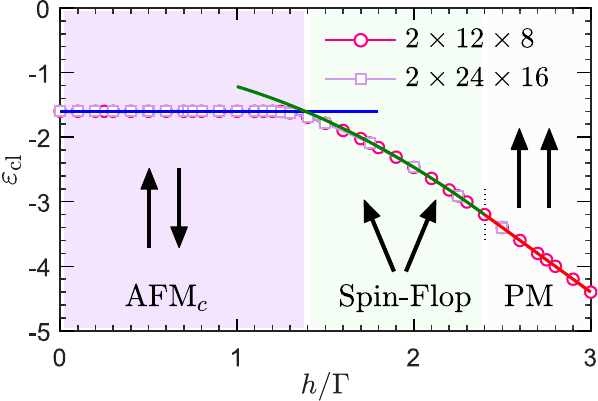}\\
\caption{Classical MC calculation of the ground-state energy $\varepsilon_{\rm cl}$ as a function of magnetic field $h/\Gamma$ $(S = 1/2)$
    in the $\Gamma$-$\Gamma'$ model with $\Gamma'/\Gamma = 0.3$.
    Two distinct clusters, $2\times12\times8$ (open circle) and $2\times24\times16$ (open square), are used in the simulation.
    The exact energy of the AFM$_c$ phase (blue line, left), spin-flop phase (green line, middle), and paramagnetic phase (red line, right) are shown for comparison.
    The cartoon patterns of these phases are shown in the inset.
    }\label{FIG-PTMC}
\end{figure}

In this subsection, we investigate the evolution of the AFM$_c$ phase under the [111] magnetic field in the region of $\Gamma, \Gamma' > 0$.
Since the applied magnetic field is parallel to the direction of the classical magnetic moment,
the energy of the AFM$_c$ phase remains unchanged in the presence of a weak magnetic field.
By contrast, a strong enough field will induce a totally polarized phase
where all the spins align along the magnetic field direction.
To quantify the value of the critical field $h_t$, we define the spin $\mathbf{S}_i$ for arbitrary $i$ as
\begin{equation}\label{EQ:U1FMHcC3}
\mathbf{S}_i = S\big(\sin\vartheta\cos\varphi \hat{\mathbf{a}} + \sin\vartheta\sin\varphi \hat{\mathbf{b}} + \cos\vartheta \hat{\mathbf{c}}\big)
\end{equation}
where $\hat{\mathbf{a}} [11\bar2]$, $\hat{\mathbf{b}} [\bar110]$, and $\hat{\mathbf{c}} [111]$ are the crystallographic axes,
$\vartheta$ and $\varphi$ are the polar angle relative to the $c$ axis and azimuthal angle in the $ab$-plane, respectively.
We note that this ansatz is certainly suitable for the paramagnetic phase, but may break down for the unpolarized phases
and thus should be checked by other methods in the intermediate region.
By using of Eq.~\eqref{EQ:U1FMHcC3}, the entire variational classical energy $e_g = E_g/NS^2$ is given by
\begin{equation}\label{EQ:FldEg}
e_g = \frac{\Gamma+2\Gamma'}{4}(1+3\cos2\vartheta) - \frac{h}{S}\cos\vartheta.  % \frac{3J+K}{2} +
\end{equation}
Strikingly, the energy is irrelevant of $\varphi$ and thus the polar angle $\vartheta$ is the sole variational parameter \cite{JanAndVoj2017}.
The optimal value $\vartheta_o$ is determined by the conditional equations
$\partial e_g/\partial\vartheta = 0$ and $\partial e_g^2/\partial\vartheta^2 > 0$,
and from which we obtain that
\begin{align}\label{EQ:PltzdTht}
\vartheta_o =
\left\{
  \begin{array}{ll}
    0,                          &   h \geq h_t \\
    \cos^{-1}\big(h/h_t\big),   &   h < h_t
  \end{array}
\right.,
\end{align}
where $h_t = 3S(\Gamma+2\Gamma')$ is the critical magnetic field.
The Eq.~\eqref{EQ:PltzdTht} indicates that just below the critical magnetic field $h_t$,
the spins deviate from the axial direction by a given angle $\vartheta_o$ and exhibit a nonzero in-plane magnetization (see the inset of Fig.~\ref{FIG-PTMC}).
With the decrease of the magnetic field, the energy of the intermediate phase grows,
and it is replaced by the AFM$_c$ phase when $h < h_t/\sqrt3$.
By substituting Eq.~\eqref{EQ:PltzdTht} into Eq.~\eqref{EQ:FldEg} and with the energy of the AFM$_c$ phase in mind,
we have the classical energy of the three phases
\begin{align}\label{EQ:ExactClEg}
e_g =
\left\{
  \begin{array}{ll}
    -\frac{3h-h_t}{3S},         &   h \geq h_t \\
    -\frac{3h^2+h_t^2}{6Sh_t},  &   h_t/\sqrt3 < h < h_t \\
    -\frac{h_t}{3S},            &   h \leq h_t/\sqrt3
  \end{array}
\right..
\end{align}
One could notice that all the three phases have two-site unit cells, see the cartoon patterns shown in Fig.~\ref{FIG-PTMC}.

\begin{figure*}[htb]
\centering
\includegraphics[width=0.75\linewidth, clip]{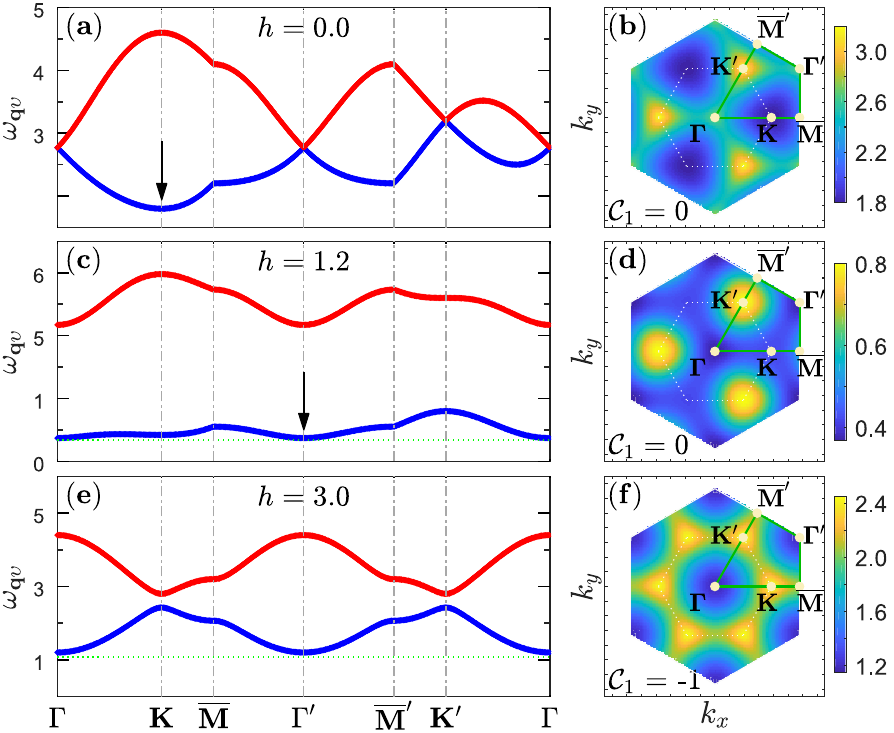}\\
\caption{Magnon dispersions $\omega_{\bm{q}1}$ (blue) and $\omega_{\bm{q}2}$ (red) of the AFM$_c$ phase or the paramagnetic phase
    in the $S = 1/2$ $\Gamma$-$\Gamma'$ model with $\Gamma'/\Gamma = 0.3$.
    (a) and (b) show the magnon dispersions and the intensity of the lowest magnon branch in the AFM$_c$ phase with $h = 0.0$, respectively.
    The black arrow in panel (a) marks the position of the minimal excitation gap, and the path in the reciprocal space is depicted in panel (b).
    (c) and (d) show the same quantities as these of (a) and (b) in the case of AFM$_c$ phase with $h = 1.2$.
    The green dotted horizontal line in panel (c) is a guide for the eye.
    (e) and (f) show the same quantities as these of (a) and (b) in the case of paramagnetic phase with $h = 3.0$.
    The green dotted horizontal line in panel (e) is a guide for the eye.
    In this paramagnetic phase, the Chern number $\mathcal{C}_1$ of the lowest magnon branch is $-1$.
    }\label{FIG-DispAFM}
\end{figure*}

Having discussed the consecutive transitions along the magnetic field,
we now perform the parallel tempering MC simulation to study the intermediate region in detail.
After identifications of the possible classical ground states,
we then perform the simulation on two cylinders of $2\times12\times8$ (open circle) and $2\times24\times16$ (open square).
The calculated energy of the AFM$_c$ phase (blue line) and the paramagnetic phase (red line) match nicely with the exact solutions
shown in Eq.~\eqref{EQ:ExactClEg}, see Fig.~\ref{FIG-PTMC}.
In the intermediate region, there are several large-unit-cell orderings and the selected configurations are shown in the Supplemental Material \cite{SuppMat}.
These results are at odds with Eq.~\eqref{EQ:U1FMHcC3} which assumes that all the spins have the same polar angle.
We note that, while the spin-flop phase (green line) is not the genuine ground state in the intermediate region,
its energy is very close to and yet slightly higher than the MC result.
This leaves the possibility open to legitimate the spin-flop phase at the quantum level.

\subsection{Spin-wave dispersions, topological magnons, and order-by-disorder mechanism}
In this subsection, we resort to the LSWT to study the magnon excitations of the underlying phases in the [111] magnetic field.
We start from the AFM$_c$ phase at the low-field region and its BdG Hamiltonian in Eq.~\eqref{EQ:H120SW} takes the form of
\begin{align}\label{EQ:BdGHam}
\hat{\vec{H}} =
	\left(\begin{array}{@{}cc|cc@{}}
	\varepsilon_0^{+}           &   \lambda_0(\bm{q})           &       0                       & \lambda_1(\bm{q})          \\
    \lambda_0^*(\bm{q})         &   \varepsilon_0^{-}           &   \lambda_1(-\bm{q})          & 0                          \\
    \hline
    0                           &   \lambda_1^*(-\bm{q})        &   \varepsilon_0^{+}           & \lambda_0^*(-\bm{q})       \\
    \lambda_1^*(\bm{q})         &   0                           &   \lambda_0(-\bm{q})          & \varepsilon_0^{-}
	\end{array}\right).
\end{align}
The momentum-dependent coupling expressions are
\begin{eqnarray}\label{EQ:BdGAFM}
  \varepsilon_0^{\pm}   &=& 2\big(\Gamma + 2\Gamma'\big) \pm {h}/{S}, \\
  \lambda_0(\bm{q})     &=& 2(\Gamma -  \Gamma') \gamma_{1,\bm{q}},  \\
  \lambda_1(\bm{q})     &=& -(\Gamma + 2\Gamma') \gamma_{0,\bm{q}}.
\end{eqnarray}
For convenience, we introduce three auxiliary functions
\begin{eqnarray}
\gamma_{0,\bm{q}} &=& \frac13\big(e^{\ii{\bm{q}\boldsymbol\delta_{x}}} + e^{\ii{\bm{q}\boldsymbol\delta_{y}}} +e^{\ii{\bm{q}\boldsymbol\delta_{z}}} \big),    \\
\gamma_{1,\bm{q}} &=& \frac13\big(\omega^{-1} e^{\ii{\bm{q}\boldsymbol\delta_{x}}} + \omega e^{\ii{\bm{q}\boldsymbol\delta_{y}}} +e^{\ii{\bm{q}\boldsymbol\delta_{z}}} \big),     \\
\gamma_{2,\bm{q}} &=& \frac13\big(\omega e^{\ii{\bm{q}\boldsymbol\delta_{x}}} + \omega^{-1} e^{\ii{\bm{q}\boldsymbol\delta_{y}}} +e^{\ii{\bm{q}\boldsymbol\delta_{z}}} \big),
\end{eqnarray}
which satisfy the relations
$\gamma^{*}_{0,\bm{q}} = \gamma_{0,-\bm{q}}$, $\gamma^{*}_{1,\bm{q}} = \gamma_{2,-\bm{q}}$, and $\gamma^{*}_{2,\bm{q}} = \gamma_{1,-\bm{q}}$.
The Berry curvature $\mathcal{F}_{\upsilon}(\bm{q})$ associated with each magnon band is given by
\begin{equation}\label{EQ:BerryCur}
\mathcal{F}_{\upsilon}(\bm{q}) = \frac{\partial \mathcal{A}^{(\upsilon)}_y(\bm{q})}{\partial q_x} - \frac{\partial \mathcal{A}^{(\upsilon)}_x(\bm{q})}{\partial q_y},
\end{equation}
where $\mathcal{A}^{(\upsilon)}_{\gamma}(\bm{q}) = \imath\textrm{Tr}\big[\mathcal{P}_{\upsilon}\boldsymbol{\Sigma}\bm{T}_{\bm{q}}^{\dagger}\boldsymbol{\Sigma}(\partial_{q_{\gamma}}\bm{T}_{\bm{q}})\big]$ ($\gamma = x, y$)
is the Berry potential.
Here, $\mathcal{P}_{\upsilon}$ is a diagonal matrix taking +1 for the $\upsilon$-th diagonal component and zero otherwise.
Alternatively, the Berry curvature can be rewritten as \cite{LuGKJ2019}
\begin{equation}\label{EQ:BerryCur-LuFormula}
\mathcal{F}_{\upsilon}(\bm{q}) = -2\textrm{Im} \bigg[\sum_{\mu\neq\upsilon}\sigma_{\mu\upsilon}
\frac{\langle\bm{T}_{\upsilon}|\partial_{q_x}\hat{\vec{H}}|\bm{T}_{\mu}\rangle \langle\bm{T}_{\mu}|\partial_{q_y}\hat{\vec{H}}|\bm{T}_{\upsilon}\rangle}
{(\omega_{\bm{q}\mu}-\omega_{\bm{q}\upsilon})^2}\bigg]
\end{equation}
with $\sigma_{\mu\upsilon} = \Sigma_{\mu\mu}\Sigma_{\upsilon\upsilon}$.
The Chern number of the $\upsilon$-th branch is obtained as the sum of the Berry curvature in the Brillouin zone,
\begin{equation}\label{EQ:Chern}
\mathcal{C}_{\upsilon} = \frac{1}{2\pi}\int_{\textrm{BZ}} \mathcal{F}_{\upsilon}(\bm{q}) d^2{\bm{q}}.
\end{equation}

Figure~\ref{FIG-DispAFM}(a) shows two magnon branches along the high-symmetry points in the Brillouin zone in the zero-field limit,
and the intensity of the lower branch in the reciprocal space is shown in Fig.~\ref{FIG-DispAFM}(b).
The $\textbf{K}$ and $\textbf{K}'$ points are inequivalent, which is reminiscent of the time-reversal symmetry breaking.
The magnon bands are gapped and the lowest excitation gap comes from the $\textbf{K}$ point with the value of
\begin{equation}\label{EQ:AFMGapK}
\Delta_{\textbf{K}} = 2(\Gamma+2\Gamma') - \sqrt{(h/S)^2 + 4(\Gamma-\Gamma')^2}.
\end{equation}
Apparently, the zero-field magnon gap $\Delta_{\textbf{K}} = 6\Gamma' = 1.8$, consistent with the data shown in Fig.~\ref{FIG-DispAFM}(a).
However, one finds that depending on the relative magnitude of the magnetic field
there could be a soft mode around the ordering wavevectors of $\boldsymbol{\Gamma}/\boldsymbol{\Gamma}'$ points.
Hence, the lowest excitation gap is given by
\begin{equation}\label{EQ:AFMGapG}
\Delta_{\boldsymbol{\Gamma}'} = \sqrt{3}(\Gamma+2\Gamma') - h/S,
\end{equation}
which decreases linearly with the increase of magnetic field.
In Fig.~\ref{FIG-DispAFM}(c), we show the magnon dispersions at a field of $h = 1.2$,
together with a lower magnon branch in Fig.~\ref{FIG-DispAFM}(d).
It is observed that excitation gap at $\boldsymbol{\Gamma}'$ point is slightly smaller than that of the $\textbf{K}$ point.
Since the melting of the AFM$_c$ ordering is accompanied by the closure of excitation gap at $\boldsymbol{\Gamma}'$ point,
the lower transition point is estimated as $h_{t1} = \sqrt3 S(\Gamma+2\Gamma')$.

We also calculate the Chern numbers and find that they are zero for both branches.
The reason may be that the two branches touch each other at some points and thus the Chern number is not well-defined.
We note that the same conclusion was drawn in a relevant study \cite{ZhuBoson2020}.
However, it is demonstrated that magnons in the paramagnetic phase is topologically nontrivial for the existence of nonzero Chern number \cite{McClarty2018,Joshi2018,LuoChen2020}.
To this end, we proceed with the analysis of the paramagnetic phase at large enough magnetic field.
Similarly, the BdG Hamiltonian of the paramagnetic phase takes the form of
\begin{eqnarray}\label{EQ:BdGAFM}
  \varepsilon_0^{\pm}  &=& -2\big(\Gamma + 2\Gamma'\big) + {h}/{S}, \\
  \lambda_0(\mathbf q) &=& -(\Gamma + 2\Gamma') \gamma_{0,\bm{q}},  \\
  \lambda_1(\mathbf q) &=& 2(\Gamma -  \Gamma') \gamma_{1,\bm{q}}.
\end{eqnarray}
The magnon spectrum at the $\boldsymbol{\Gamma}$ point is
\begin{equation}\label{EQ:PlrzGapG}
\Delta_{\boldsymbol{\Gamma}} = h/S - 3(\Gamma+2\Gamma'),
\end{equation}
which increases linearly with the magnetic field when $h$ is larger than the upper transition point $h_{t2} = 3S(\Gamma+2\Gamma')$.
Figure~\ref{FIG-DispAFM}(e) shows the dispersion of the paramagnetic phase where $h = 3.0$ is taken as an example.
It can be verified that the magnon gap at $\boldsymbol{\Gamma}$ is 1.2, which is in accordance with the theoretical value revealed in Eq.~\eqref{EQ:PlrzGapG}.
More importantly, our result suggests that the Chern numbers of the two branches in the paramagnetic phase are $-1$ and $+1$, respectively, see Fig.~\ref{FIG-DispAFM}(f).

\begin{figure}[!ht]
\centering
\includegraphics[width=0.95\columnwidth, clip]{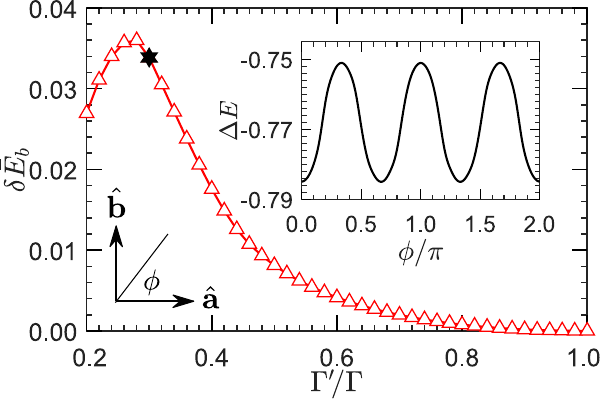}\\
\caption{Energy barrier $\delta\bar{E}_b$ between the spin-flop phase of different orientations along the line of $h = 9(\Gamma+2\Gamma')/8$.
    Inset: Spin-wave energy correction $\Delta E(\varphi)$ at the parameter point $(\Gamma'/\Gamma, h) = (0.3, 1.8)$ vs azimuthal angle $\varphi$,
    which is suited at the $\hat{\textbf{a}}$-$\hat{\textbf{b}}$ plane.
    }\label{FIG-ObDNmtc}
\end{figure}

The LSWT analysis shows clearly that there should be an intermediate region in the window of $h \in (h_{t1}, h_{t2})$,
which happens to be the same interval inferred from the classical study (see Eq.~\eqref{EQ:ExactClEg}).
For the spin-flop phase, the classical moment direction $\textbf{S}_i(\vartheta, \varphi)$ is shown in Eq.~\eqref{EQ:U1FMHcC3}
where $\vartheta$ is given by Eq.~\eqref{EQ:PltzdTht}.
According to Eq.~\eqref{EQ:SpinWaveEg} we calculate the spin-wave energy $\varepsilon_{\rm sw}$ at a magnetic field of $h = 1.8$,
and the energy reduction $\Delta E(\varphi) = \varepsilon_{\rm sw} - S^2\varepsilon_{\rm cl}$ with respect to the azimuthal angle $\varphi$
is shown in the inset of Fig.~\ref{FIG-ObDNmtc}.
It is shown that $\Delta E(\varphi)$ exhibits a period of $2\pi/3$ in the in-plane manifold
and the angles at $\varphi$ = 0, $2\pi/3$, and $4\pi/3$ are more energetically favorable than the others.
Hence, the emergent classical $U(1)$ manifold is lifted by quantum fluctuations, generating a discrete $C_3$ rotational symmetry.
We also introduce the energy barrier $\delta\bar{E}_b$ as $E_g(\varphi = \pi) - E_g(\varphi = 0)$,
which is 0.0338 when $(\Gamma'/\Gamma, h) = (0.3, 1.8)$ (marked by a hexagram in Fig.~\ref{FIG-ObDNmtc}).
In the main panel of Fig.~\ref{FIG-ObDNmtc},
we present the energy barrier $\delta\bar{E}_b$ along the line of $h = 9(\Gamma+2\Gamma')/8$.
The value of $\delta\bar{E}_b$ gradually raises with the increase of $\Gamma'/\Gamma$ up to $\Gamma'/\Gamma \approx 0.27$.
Afterwards, it drops rapidly and vanishes at $\Gamma'/\Gamma = 1$ where the system possesses a hidden $U(1)$ symmetry.

\begin{figure}[!ht]
\centering
\includegraphics[width=0.95\columnwidth, clip]{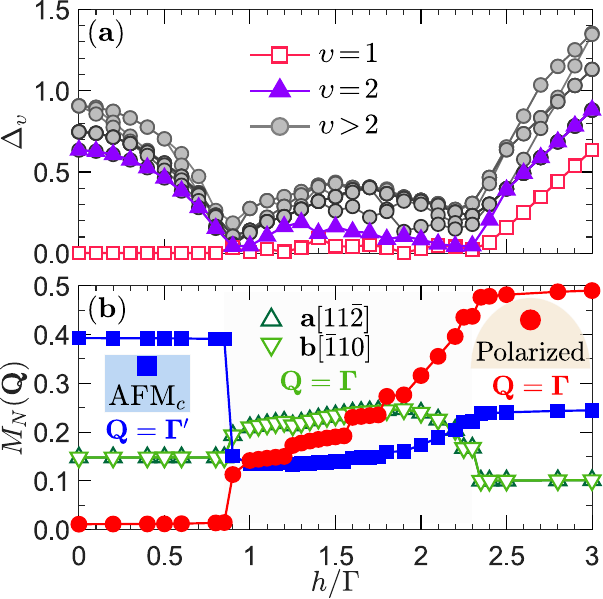}\\
\caption{(a) The first fifteen excitation gaps $\Delta_{\upsilon}$ ($\upsilon$ = 1--15) as functions of magnetic field $h/\Gamma$
    in the $\Gamma$-$\Gamma'$ model with $\Gamma'/\Gamma = 0.3$.
    (b) Magnetic order parameters $M_N(\textbf{Q})$ for the AFM$_c$ phase and paramagnetic phase with $\textbf{Q}$ = $\boldsymbol{\Gamma}'$ (blue square)
    and $\boldsymbol{\Gamma}$ (red circle), respectively.
    The intermediate spin-flop phase has a finite in-plane magnetization (green triangles).
    }\label{FIG-U1FM}
\end{figure}

\subsection{DMRG calculation}

In the preceding subsection, we predict that a spin-flop phase can occur in a wide field region before entering into the paramagnetic phase.
Here, we confirm the existence of such phase by the DMRG method.
Figure~\ref{FIG-U1FM}(a) shows the first fifteen low-lying excitation gaps $\Delta_{\upsilon} = E_{\upsilon} - E_0$
in the $\Gamma$-$\Gamma'$ model with $\Gamma'/\Gamma = 0.3$ fixed.
The method to target the first few low-lying energy levels simultaneously is shown in Ref.~\onlinecite{LuoNPJ2021}.
With the increase of the magnetic field,
the excitation gap $\Delta_2$ of the AFM$_c$ phase goes down gradually and is vanishingly small at $h_{t1} = 0.88(2)$.
Beyond the transition point, excitation gaps are small and the spectrum is very dense in a large interval, indicative of a gapless region.
Exceeding $h_{t2} = 2.38(2)$, excitation gap $\Delta_1$ opens linearly with the magnetic field.
We note in passing that the transition points are fairly consistent with those obtained by cylinder DMRG calculation \cite{SuppMat}.
In addition, magnetic order parameters of the AFM$_c$ phase and the paramagnetic phase are shown in Fig.~\ref{FIG-U1FM}(b).
For the AFM$_c$ phase, the SSF peaks at the $\boldsymbol{\Gamma}'$ point,
and the order parameter $M(\boldsymbol{\Gamma}')$ has a sharp jump at $h_{t1}$,
signifying a first-order transition thereof.
In the intermediate phase, the spins are only partially polarized as opposed to the paramagnetic phase when $h_t > h_{t2}$.
However, a nontrivial observation is that it also has a \textit{uniform} in-plane correlation that is perpendicular to the external field.
For example, magnetization along $\hat{\textbf{a}}$ and $\hat{\textbf{b}}$
are of equal strength and are overlapped in the plot (see Fig.~\ref{FIG-U1FM}(c)).
Consequently, the intermediate phase is recognized as a gapless spin-flop phase with a temporarily emergent $U(1)$ symmetry.
The finite-size scaling of the order parameters and the SSF of the spin-flop phase are shown in the Supplemental Material \cite{SuppMat}.

\begin{figure}[!ht]
\centering
\includegraphics[width=0.95\columnwidth, clip]{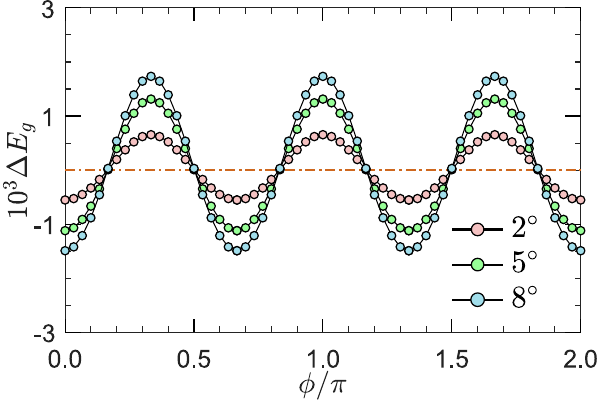}\\
\caption{Behaviors of the shifted ground-state energy $\Delta E_g$ in the range of $\varphi \in [0, 2\pi)$ for three different tilted angle $\vartheta$ =
    $2^{\circ}$ (pink), $5^{\circ}$ (green), and $8^{\circ}$ (blue) in the $\Gamma$-$\Gamma'$ model under a $c$-axis magnetic field.
    Here, the DMRG calculation is performed on a 24-site cluster with ($\Gamma'/\Gamma$, $h$) = (0.3, 1.8).
    The energy valleys at $\varphi$ = 0, $2\pi/3$, and $4\pi/3$ are reminiscent of the order-by-disorder phenomenon.
    }\label{FIG-ObDU1FM}
\end{figure}

However, the in-plane component of the spin-flop phase is likely unstable against extra perturbation.
The emergent $U(1)$ symmetry is then broken down to $C_3$ rotational symmetry,
accompanied by the appearance of gapless Goldstone modes.
To this end, we apply a tilted magnetic field $\textbf{h} = h(\vartheta,\varphi)$ which enjoys the same form of Eq.~\eqref{EQ:U1FMHcC3}.
Here, the intensity of the field $h = 1.8$ and the tilted angle $\vartheta$ relative to the $c$-axis is specified as $2^{\circ}$, $5^{\circ}$, and $8^{\circ}$.
Figure~\ref{FIG-ObDU1FM} shows the behaviors of the shifted ground-state energy $\Delta E_g$,
which is defined as $E_g - \bar{E}_g$ with $\bar{E}_g = \frac{1}{2\pi}\int_0^{2\pi}d\varphi E_g(\varphi)$,
with respect to the in-plane azimuthal angle $\varphi \in [0, 2\pi)$ \cite{GohlkeCKK2020}.
It can be observed that the variation of the energy is one order of magnitude smaller than that of the semiclassical situation.
However, in both cases there is a breaking of the continuous $U(1)$ symmetry to the discrete $C_3$ rotational symmetry,
giving rise to three local minima when $\varphi$ = 0, $2\pi/3$, and $4\pi/3$.
In addition, the energy barrier $\delta\bar{E}_b$ = $\Delta E_g(\varphi = \pi)$ $-$ $\Delta E_g(\varphi = 0)$ obeys approximately the fitting formula
$\delta\bar{E}_b \approx 10^{-3}\cdot\frac{7.1\vartheta}{9.6+\vartheta}$,
showing that the energy barrier $\delta\bar{E}_b$ will be less sensitive to the tilted angle $\vartheta$ as $\vartheta$ increases.
To conclude, there is a two-step symmetry changing in the spin-flop phase .
The first step is from the discrete symmetry to the emergent $U(1)$ symmetry, while the second step is from $U(1)$ symmetry to the broken $C_3$ rotational symmetry.
We note that a similar phenomenon is also reported in the classical honeycomb $\Gamma$ model in a magnetic field \cite{Tian2021}.

As shown in Fig.~\ref{FIG-FldhcPD}(b), the spin-flop phase could survive for at least $\Gamma'/\Gamma = 1$,
at which the model is equivalent to the spin-$1/2$ XXZ model in a longitudinal magnetic field with an easy-axis anisotropy $\Delta_a = 2$.
Accidentally, the spin-flop phase could also be interpreted as the superfluid phase
in the extended Bose-Hubbard model whose Hamiltonian reads \cite{Wessel2007,GanWenYeetal2007}
\begin{align}\label{EQ:ExtBH-Ham}
\mathcal{H}_{\textrm{BH}} =
    & -t \sum_{\left<ij\right>}(b_i^{\dagger}b_j + b_ib_j^{\dagger}) + V\sum_{\left<ij\right>} n_in_j - \mu\sum_i n_i   \nonumber \\
    & + \frac{U}{2}\sum_i n_i(n_i-1),
\end{align}
where $b_i^{\dagger}$ ($b_i$) is the creation (annihilation) operator at site $i$
and $n_i = b_i^{\dagger}b_i$ is the corresponding occupation number.
Here, $t$ is the nearest-neighbor hopping parameter, $\mu$ is the chemical potential,
and $U$ and $V$ represent the on-site and nearest-neighbor repulsive interactions, respectively.
In the hard-core limit where $U/t \to \infty$, there is one boson at most on each site.
By virtue of the mapping $b_i^{\dagger} = S_i^+$, $b_i = S_i^-$, and $n_i = S_i^z + 1/2$,
Eq.~\eqref{EQ:ExtBH-Ham} can be mapped onto the spin-$1/2$ XXZ model under a longitudinal magnetic field,
\begin{align}\label{EQ:XXZFld-Ham}
\overline{\mathcal{H}} =
    & 2t\Big[\sum_{\left<ij\right>}\big[-(S_i^xS_j^x \!+\! S_i^yS_j^y) \!+\! \Delta_{a} S_i^zS_j^z\big] \!-\! h\sum_i S_i^z \Big] \nonumber\\
    & + E_0,
\end{align}
where $\Delta_a = V/2t$ is the anisotropy of the spin-spin interaction,
$h = \frac{1}{2t}\big(\mu-zV/2\big)$ = $\Delta\big(\mu/V-z/2\big)$ is the longitudinal magnetic field,
and $E_0 = -\frac{NV}{2}\big(\mu/V-z/4\big)$ is an energy constant
arising from the mapping between the spins and bosons operators.
Considering the case $t/V = 1/4$ (inversely, we have $\Delta_a = V/2t = 2$)
and $\mu/V > z/2 = 1.5$ in the original extended Bose-Hubbard model,
the ground state is a solid with density $\rho = 1/2$ when $\mu/V$ is sightly increased,
a $\rho = 1$ Mott insulator at large enough $\mu/V$, and a superfluid at moderate $\mu/V$.
In view of the relation $\mu_t/V = (h_t + z)/2$ with $h_{t,1} = 2.115(3)$ and $h_{t,2} = 4.5$,
our result suggests the first transitions occurs at $\mu_{t,1}/V = 2.558(2)$ and $\mu_{t,2}/V = 3.75$,
which is fairly consistent with quantum Monte Carlo simulations
(for illustration, see Ref.~[\onlinecite{Wessel2007,GanWenYeetal2007}] and also Supplemental Material \cite{SuppMat}).

%%%%%%%%%%%%%%%%%%%%%%%%%%%%%%%%%%%%%%%%%%%%%%%%%%%%%%%%%%%%%%%%%%%%%%%%%%%%%%
\section{Conclusion}\label{SEC:CONC}
%%%%%%%%%%%%%%%%%%%%%%%%%%%%%%%%%%%%%%%%%%%%%%%%%%%%%%%%%%%%%%%%%%%%%%%%%%%%%%

In this paper we focus on the interplay of magnetic field and trigonal distortion \cite{RauKeeArXiv2014} in honeycomb $\Gamma$ model.
For this purpose, we have studied a $\Gamma$-$\Gamma'$ model in a [111] magnetic field in the vicinity of a dominated AFM $\Gamma$ region.
In the absence of magnetic field, a 120$^{\circ}$ phase and an AFM$_c$ phase can be selected immediately from the infinitely degenerate ground state
of the classical $\Gamma$ model, depending on the sign of $\Gamma'$ interaction.
The classical 120$^{\circ}$ phase is unstable against quantum fluctuations, giving away to the neighboring zigzag ordering.
At the quantum level, two exotic phases are found to exist in the intermediate region between the zigzag phase and the AFM$_c$ phase.
One is a $\Gamma$SL stemming from the ground state of honeycomb $\Gamma$ model,
while the other is a chiral spin state which spontaneously breaks the time-reversal symmetry.
Upon applying a magnetic field, regions of the $\Gamma$SL and chiral spin state are enlarged but are terminated
before entering the paramagnetic phase at large field.

A nontrivial observation of this work is that,
there is a field-induced spin-flop phase as long as a modest magnetic field is applied over the AFM$_c$ phase.
The spins in the spin-flop phase are tilted away from the field direction and are free to rotate in the honeycomb plane,
indicative of an emergent $U(1)$ symmetry.
Due to the quantum fluctuation in the frustrated magnet, such a continuous symmetry is broken down to the $C_3$ rotational symmetry
where the spins are perpendicular to any of the three types of bonds.
When $\Gamma'/\Gamma = 1$, the model is reduced to an easy-axis spin-$1/2$ XXZ antiferromagnet subjected to a longitudinal magnetic field.
In this circumstance, it is equivalent to a hard-core extended Bose-Hubbard model.
In that sense, the spin-flop phase is merely the superfluid phase.
In doing so, we manifest an unusual route from the $\Gamma$ region to the XXZ magnet.

In closing, we comment that there are several ways to achieve such a dominated $\Gamma$ interaction in experiments.
In $\alpha$-RuCl$_3$, for example, the spin interactions are revealed to be sensitive to the layer stacking and octahedral distortion,
and the overwhelming $\Gamma$ regime with a desired AFM $\Gamma'$ interaction could be achieved upon applying compression \cite{PeterDFT2020}.
On the other hand, by virtue of the circularly-polarized light,
the Heisenberg interaction in $\alpha$-RuCl$_3$ can be made much smaller than the anisotropic exchange interactions $K$ and $\Gamma$ \cite{Arakawa2021},
and the tailored light pulse can further weaken the Kitaev interaction by a proper adjustment of its amplitude and frequency \cite{Sriram2021,Strobel2021}.
Therefore, these procedures allow us to drive the material into a regime where the $\Gamma$ interaction is prominent \cite{Kumar2021}.

%%%%%%%%%%%%%%%%%%%%%%%%%%%%%%%%%%%%%%%%%%%%%%%%%%%%%%%%%%%%%%%%%%%%%%%%%%%%%%
\begin{acknowledgments}
We would like to thank Y. Lu, X. Wang, T. Ying, J. Zhao, and Z. Zhou for useful discussions,
and are extremely grateful to J. S. Gordon and P. P. Stavropoulos for the intimate collaboration of a related research \cite{LuoStaKee2022}.
Q.L. was supported by the Fundamental Research Funds for the Central Universities (Grant No. 1018-XAA22046)
and the startup Fund of Nanjing University of Aeronautics and Astronautics (Grant No. YAH21129).
H.-Y.K. was supported by the NSERC Discovery Grant No. 06089-2016,
the Centre for Quantum Materials at the University of Toronto,
the Canadian Institute for Advanced Research, and the Canada Research Chairs Program.
Computations were performed on the Niagara supercomputer at the SciNet HPC Consortium.
SciNet is funded by: the Canada Foundation for Innovation under the auspices of Compute Canada;
the Government of Ontario; Ontario Research Fund - Research Excellence; and the University of Toronto.
\end{acknowledgments}

%%%%%%%%%%%%%%%%%%%%%%%%%%%%%%%%%%%%%%%%%%%%%%%%%%%%%%%%%%%%%%%%%%%%%%%%%%%%%%%

% \bibliography{manuscript_GammaQSLRuCl3.bib}
%%%%%%%%%%%%%%%%%%%%%%%%%%%%%%%%%%%%%%%%%%%%%%%%%%%%%%%%%%%%%%%%%%%%%%%%%%

% \bibliography{manuscript}
%

%%%%%%%%%%%%%%%%%%%%%%%%%%%%%%%%%%%%%%%%%%%%%%%%%%%%%%%%%%%%%%%%%%%%%%%%%%%%%%%%%%%%%%%%%%%%%%%%%%%%%%%%%%%%%%%%%%%%%%%%%%%%%%%%%%%%%%%%%%%%%%%%
%%%%%%%%%%%%%%%%%%%%%%%%%%%%%%%%%%%%%%%%%%%%%%%%%%%%%%%%%%%%%%%%%%%%%%%%%%%%%%%%%%%%%%%%%%%%%%%%%%%%%%%%%%%%%%%%%%%%%%%%%%%%%%%%%%%%%%%%%%%%%%%%

%%%%%%%%%% Merge with supplemental materials %%%%%%%%%%
%% ref: https://arxiv.org/abs/1709.10096

%%%%%%%%%%%%%%%%%%%%%%%%%%%%%%%%%%%%%%%%%%%%%%%%%%%%%%%%%%%%%%%%%%%%%%%%%%%%%%%%%%%%%%%%%%%%%%%%%%%%%%%%%%%%%%%%%%%%%%%%%%%%%%%%%%%%%%%%%%%%%%%%

\clearpage

\onecolumngrid

%%%%%%%%%% Merge with supplemental materials %%%%%%%%%%
%%%%%%%%%% Prefix a "S" to all equations, figures, tables and reset the counter %%%%%%%%%%
\newpage

\newcounter{sectionSM}
\newcounter{equationSM}
\newcounter{figureSM}
\newcounter{tableSM}
\stepcounter{equationSM}
\setcounter{section}{0}
\setcounter{equation}{0}
\setcounter{figure}{0}
\setcounter{table}{0}
\setcounter{page}{1}
\makeatletter
\renewcommand{\thesection}{\textsc{S}\arabic{section}}
\renewcommand{\theequation}{\textsc{S}\arabic{equation}}
\renewcommand{\thefigure}{\textsc{S}\arabic{figure}}
\renewcommand{\thetable}{\textsc{S}\arabic{table}}

% \onecolumngrid

%%%%%%%%%%%%%%%%%%%%%%%%%%%%%
\begin{center}
{\large{\bf Supplemental Material for\\
 ``Interplay of magnetic field and trigonal distortion in honeycomb $\Gamma$ model:\\ Occurrence of a spin-flop phase''}}
\end{center}
\begin{center}
Qiang Luo$^{1,\,2}$ and Hae-Young Kee$^{1,\,3}$\\
\quad\\
$^1$\textit{Department of Physics, University of Toronto, Toronto, Ontario M5S 1A7, Canada}\\
$^2$\textit{College of Physics, Nanjing University of Aeronautics and Astronautics, Nanjing, 211106, China}\\
$^3$\textit{Canadian Institute for Advanced Research, Toronto, Ontario, M5G 1Z8, Canada}
\end{center}

\twocolumngrid

%\begin{widetext}
%\end{widetext}
%\widetext

%%%%%%%%%%%%%%%%%%%%%%%%%%%%%
%In this supplemental material~(SM), we collect a detailed derivation of the fidelity susceptibility~(FS) presented in the main text.
%For simplicity, we will still assume
%$|\gamma|<1$ and $h^2+\gamma^2>1$ so as to avoid the imaginary unit $i$.

\section{From the $\Gamma$-$\Gamma'$ model to the extended Bose-Hubbard model}

\subsection{Mapping to the XXZ model}

The Hamiltonian shown in Eq.~(\textcolor{red}{1}) in the main text could have two different frames.
One is the cubic $xyz$ basis and the Hamiltonian takes the well-known JK$\Gamma\Gamma'$ form,
while the other is the  crystallographic $abc$ reference frame where the perpendicular direction is the ${\textbf{c}}$ axis.
Mathematically, we have ${\textbf{a}} [11\bar2]$, ${\textbf{b}} [\bar110]$, and ${\textbf{c}} [111]$.
We can replace the cubic spin components $\left(S_i^x, S_i^y, S_i^z\right)^T$ by
\begin{eqnarray}\label{RotaCubic}
\left(\begin{array}{ccc}
    \frac{1}{\sqrt6}        &   - \frac{1}{\sqrt2}     &    \frac{1}{\sqrt3}        \\
    \frac{1}{\sqrt6}        &   \frac{1}{\sqrt2}       &    \frac{1}{\sqrt3}        \\
    -\sqrt{\frac{2}{3}}     &   0                      &    \frac{1}{\sqrt3}        \\
\end{array}\right)
\left(\begin{array}{c}
    \tilde{S}_i^x \cos\varphi + \tilde{S}_i^y \sin\varphi       \\
    -\tilde{S}_i^x \sin\varphi + \tilde{S}_i^y \cos\varphi       \\
    \tilde{S}_i^z        \\
\end{array}\right),
\end{eqnarray}
and the Hamiltonian is then casted into \cite{SMChalKhal2015,SMMaksimovCherny2020,SMLuoChen2020}
\begin{align}\label{JzzJppmm-Ham}
\mathcal{H} =& \sum_{\left<ij\right>} \Big[{J_{zz}\tilde{S}_i^z\tilde{S}_j^z+J_{\pm}(\tilde{S}_i^+\tilde{S}_j^- + \tilde{S}_i^-\tilde{S}_j^+)} \nonumber\\
             & +{J_{\pm\pm}(\gamma_{ij}\tilde{S}_i^+\tilde{S}_j^+ + \gamma_{ij}^{*}\tilde{S}_i^-\tilde{S}_j^-)}  \nonumber\\
             & J_{z\pm}(\gamma_{ij}^{*}\tilde{S}_i^+ \tilde{S}_j^z + \gamma_{ij}\tilde{S}_i^-\tilde{S}_j^z + \langle i \leftrightarrow j\rangle)\Big]
\end{align}
where $\tilde{S}_{i}^{\alpha}$ ($\alpha=x,y,z$) are the three components of spin-1/2 operators,
and $\tilde{S}_{i}^{\pm}=\tilde{S}_i^x \pm i\tilde{S}_i^y$.
The phase factor $\gamma_{ij} = e^{\imath \varphi_{ij}}$
where $\varphi_{ij} = 2\pi/3, -2\pi/3$, and 0 for the bonds $\langle{ij}\rangle$
along the $\textbf{X}$, $\textbf{Y}$, and $\textbf{Z}$ direction, respectively.
The interaction parameters are
\begin{equation}\label{IntParasTransJKGGp}
\begin{split}
& J_{zz}= J + \frac{K+2(\Gamma+2\Gamma')}{3}, \\
& J_{\pm}= \frac{J}{2} + \frac{K-(\Gamma+2\Gamma')}{6}, \\
& J_{\pm\pm}= \frac{K+2(\Gamma-\Gamma')}{6}, \\
& J_{z\pm}= \frac{-K + (\Gamma-\Gamma')}{3\sqrt2}.
\end{split}
\end{equation}
We emphasize that the energy spectra of Eq.~\eqref{JzzJppmm-Ham} are invariant if we alter the sign of $J_{z\pm}$.
This can be elucidated by a $\pi$-rotation around the $z$ axis in the spin space, i.e.,
$\tilde{S}_i^z \to \tilde{S}_i^z$ and $\tilde{S}_i^{\pm} \to -\tilde{S}_i^{\pm}$,
resulting in $J_{z\pm} \to -J_{z\pm}$, while other couplings remain unchanged.

For the $\Gamma$-$\Gamma'$ model with equal strength of $\Gamma$ and $\Gamma'$,
$J_{\pm\pm}$ and $J_{z\pm}$ vanish and Eq.~\eqref{JzzJppmm-Ham} reduces to
\begin{align}\label{XXZG-Ham}
\mathcal{H} =& \Gamma \sum_{\left<ij\right>} \Big[-\frac12(\tilde{S}_i^+\tilde{S}_j^- + \tilde{S}_i^-\tilde{S}_j^+) + 2\tilde{S}_i^z\tilde{S}_j^z\Big],
\end{align}
which is nothing but the \textit{easy-axis} XXZ model with a $\mathbb{Z}_2 \ltimes U(1)$ symmetry.
% Specifically, if $\Gamma > 0$, the spins on the two sublattices point along $+\tilde{z}$ and $-\tilde{z}$ directions, respectively.

% \vspace{-0.20cm}
\subsection{Bose-Hubbard model in hard-core limit}

\begin{figure}[!ht]
\centering
\includegraphics[width=0.95\columnwidth, clip]{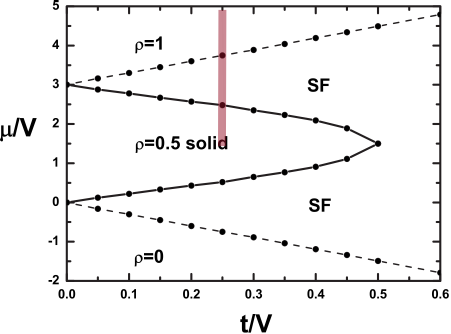}\\
\caption{Quantum phase diagram of the hard-core Bose-Hubbard model on a honeycomb lattice \cite{SMGanWenYeetal2007}.
  Along the vertical line where $t/V = 0.25$, transition between $\rho=0.5$ solid and superfluid occurs at $\mu_{t,l}/V \approx 2.5$,
  while transition between $\rho=1$ insulator and superfluid occurs at $\mu_{t,l}/V \approx 3.75$.
  }\label{FIGSM-BHSuperfluid}
\end{figure}

In the hard-core limit where $U/t \to \infty$, the extended Bose-Hubbard model is equivalent to the spin-$1/2$ XXZ model under a longitudinal magnetic field.
For the latter model, the easy-axis anisotropy $\Delta_a = V/2t$
and the effective magnetic field $h = \frac{1}{2t}\big(\mu-zV/2\big)$ = $\Delta_a \big(\mu/V-z/2\big)$.
Here, $z = 3$ is the coordination number of the honeycomb lattice.
When $\Gamma' = \Gamma = 1$,
it is shown in Eq.~\eqref{XXZG-Ham} that the anisotropy $\Delta_a = 2$ in the $\Gamma$-$\Gamma'$ model.
Also, the transition points of the AFM$_c$--spin-flop transition and the paramangetic--spin-flop transition are 2.115(3) and 4.5
(see Fig.~\textcolor{red}{3}(b) in the main text), respectively.
In view of the relation $\mu_t/V = (h_t + 3)/2$, we have $\mu_{t,l}/V = 2.558(2)$ and $\mu_{t,h}/V = 3.75$,
which is fairly consistent with quantum Monte Carlo simulations~(see Fig.~\ref{FIGSM-BHSuperfluid}) \cite{SMGanWenYeetal2007}.

\onecolumngrid

\section{Classical $\Gamma$-$\Gamma'$ model}
In the Sec.~\textcolor{red}{V} in the main text, it is shown that starting from the AFM$_c$ phase at the zero field,
there is an intermediate region when the field is roughly in the interval of $[h_t/\sqrt3, h_t]$ with $h_t = 3S(\Gamma+2\Gamma')$.
In this region, the spin configurations are very complicated and could vary with the increase of the magnetic field.
Nevertheless, the energy of the spin-flop phase is very close to the authentic phases, and the discrepancy becomes less pronounced with increasing $h$.

\begin{figure}[!ht]
\centering
\includegraphics[width=0.75\columnwidth, clip]{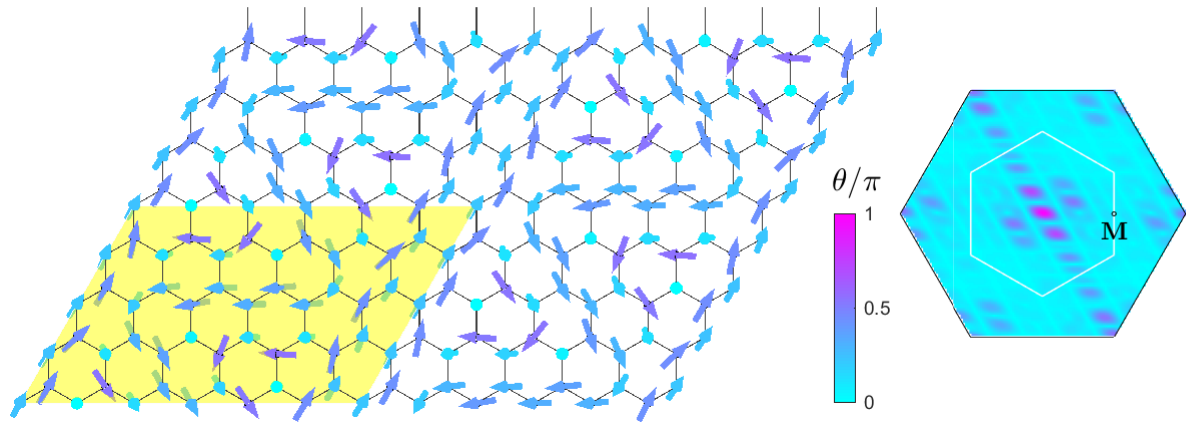}\\
\caption{Spin texture at the parameter ($\Gamma'/\Gamma$ = 0.3, $h$ = 1.5).
  The spins are parameterized by $\textbf{S}_i$ = $S (\sin\theta \cos\phi$, $\sin\theta \sin\phi$, $\cos\theta)$.
  Here, $\theta$ is represented by the color (see colormap) while $\phi$ is represented by the orientation of the arrow in the plane.
  The shape of the unit cell (with 48 sites in total) is marked by the yellow shadow.
  The right panel is the static structure factor in the momentum space.
  }\label{FIGSM-SpinTex300}
\end{figure}

\begin{figure}[!ht]
\centering
\includegraphics[width=0.75\columnwidth, clip]{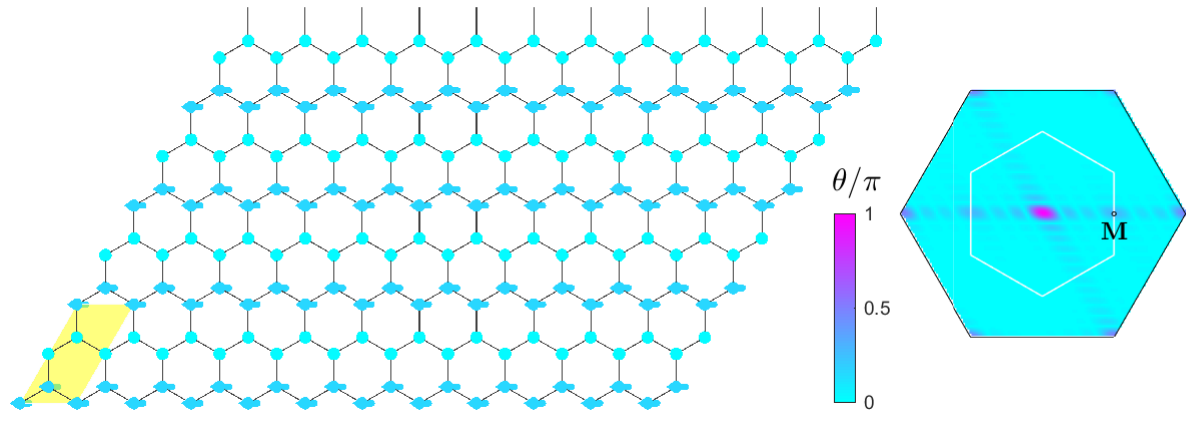}\\
\caption{The same as Fig.~\ref{FIGSM-SpinTex300} but with the parameter ($\Gamma'/\Gamma$ = 0.3, $h$ = 2.0).
  }\label{FIGSM-SpinTex400}
\end{figure}

For clarification, we focus on the line of $\Gamma'/\Gamma$ = 0.3 and the intermediate region exists in the interval of [1.3856, 2.40].
When the field is $1.50$, the spin configuration of the underlying phase is shown in Fig.~\ref{FIGSM-SpinTex300}, whose unit cell contains 48 sites.
The ground-state energy per site is 0.075 lower than that of the spin-flop phase.
However, when the field is $2.0$, the unit cell of the underlying phase is only 4 (see Fig.~\ref{FIGSM-SpinTex400}), and the energy difference is less than 0.004.

\section{Quantum $\Gamma$-$\Gamma'$ model: various calculation under periodic boundary condition}

\subsection{mapping a 24-site cluster to a one-dimensional chain}
\begin{figure}[!ht]
\centering
\includegraphics[width=0.75\columnwidth, clip]{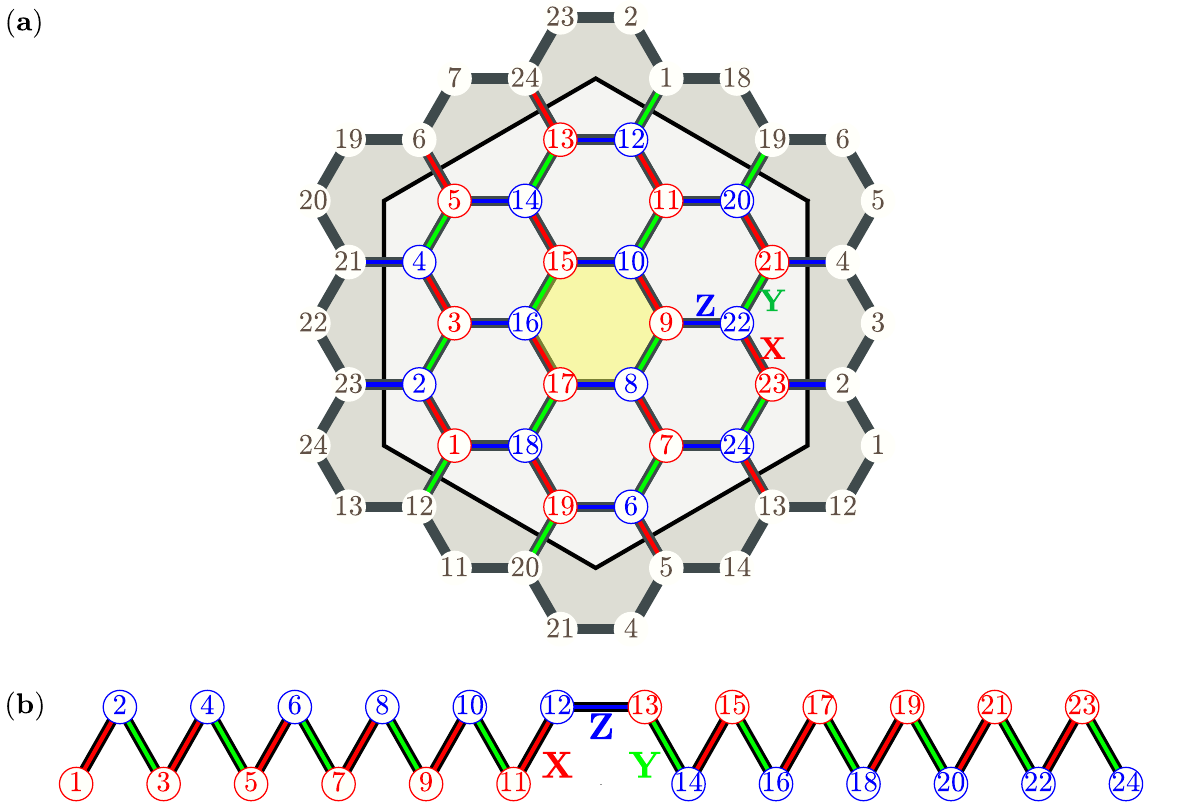}\\
\caption{(a) The numbering of the sites on the 24-site hexagonal cluster. (b) The one-dimensional chain analog of the hexagonal cluster.
  }\label{FIGSM-24toChn}
\end{figure}

The DMRG method is known as a powerful method for solving problems in one dimension.
To apply it to two-dimensional problems, one need to map the lattice geometries to snake-like chains.
Figure~\ref{FIGSM-24toChn} shows the method that maps a 24-site cluster to a one-dimensional chain.
One needs to number the sites from 1 to 24, and the sites belonged to the $A$ ($B$) sublattice are assigned odd (even) numbers.
The effective spin chain is created by connecting the sites in order.
Nevertheless, the procedure will inevitably involve long-range correlation and entanglement.
For example, in the case of the hexagonal cluster, the sites 15 and 10 are of nearest neighbor.
while site 15 is the \textit{fifth} nearest neighbor of the site 10 in the resulting spin chain.
However, all the interactions are kept in the spin chain and the full Hamiltonian remains unchanged.
Hence, the true ground state could still be targeted properly.
Since the long-range interactions are brought in,
the DMRG method is less efficient and one need to increase the number of block states and/or sweep times gradually to obtain a reliable result.
In practice, for a 24-site cluster, 3000 block states and 12 sweep times are enough to reach an energy precision with 7~8 digits after the decimal point.

\subsection{multiple transitions across the phase diagram}

\begin{figure}[!ht]
\centering
\includegraphics[width=0.45\columnwidth, clip]{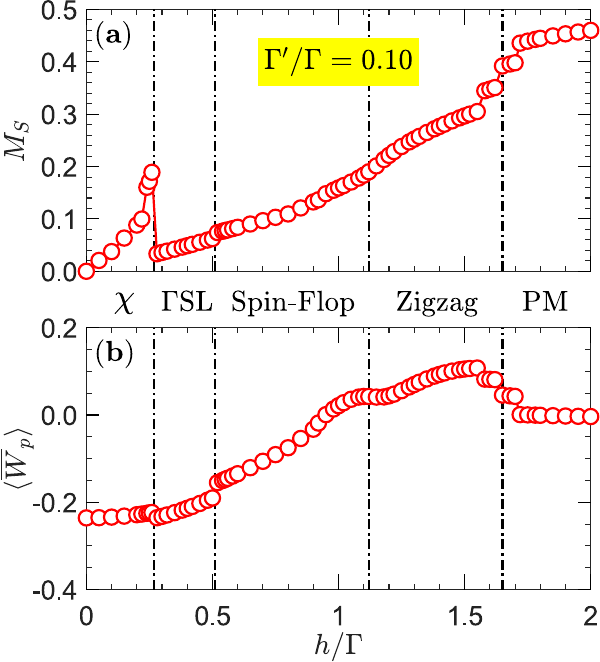}\\
\caption{(a) Behavior of the magnetization $M_S$ as a function of magnetic field $h$ in the $\Gamma$-$\Gamma'$ model with $\Gamma'/\Gamma$ = 0.1.
  The underlying geometry is a 24-site hexagonal cluster. The dash dotted lines represent the phase boundaries of the adjacent phases.
  (b) The same as (a) but for the flux-like plaquette $\langle\overline{W}_p\rangle$.
  }\label{FIGSM-Gp010}
\end{figure}

In this subsection we will show the method to determine the phase transitions in the phase diagram.
Figure~\ref{FIGSM-Gp010} presents the results along the line of $\Gamma'/\Gamma$ = 0.1.
Figure~\ref{FIGSM-Gp010}(a) shows the magnetization $M_S=\sqrt{\left\langle S_x\right\rangle^2+\left\langle S_y\right\rangle^2+\left\langle S_z\right\rangle^2}$,
which undergoes five distinct phases termed as chiral-spin ($\chi$) phase, $\Gamma$SL, spin-flop phase, zigzag phase, and paramagnetic phase.
Similarly, the flux-like plaquette $\langle{\overline{W}}_p\rangle$ shown in Fig.~\ref{FIGSM-Gp010}(b) displays the same phase transitions,
corroborating the reliability of our phase diagram.

\begin{figure}[!ht]
\centering
\includegraphics[width=0.45\columnwidth, clip]{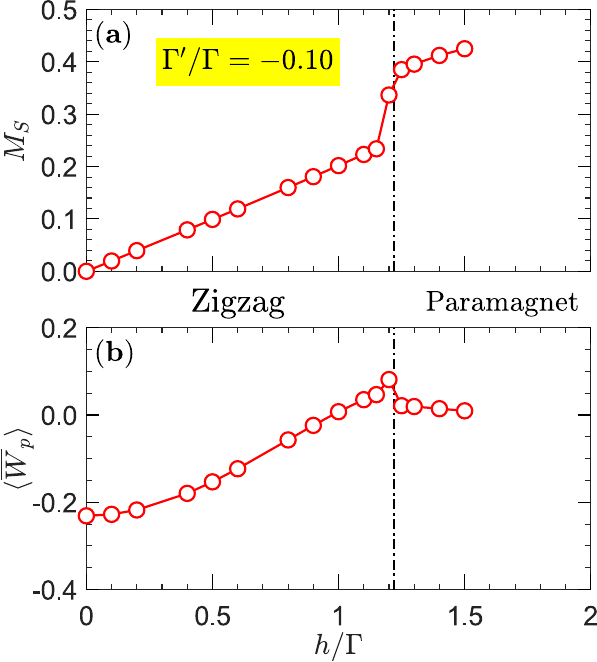}\\
\caption{(a) Behavior of the magnetization $M_S$ as a function of magnetic field h in the $\Gamma$-$\Gamma'$ model with $\Gamma'/\Gamma$ = $-0.1$.
  The underlying geometry is a 24-site hexagonal cluster. The dash dotted lines represent the phase boundaries of the adjacent phases.
  (b) The same as (a) but for the flux-like plaquette $\langle\overline{W}_p\rangle$.
  }\label{FIGSM-Gpn010}
\end{figure}

We also show the phase transition along the line of $\Gamma'/\Gamma = -0.1$ in Fig.~\ref{FIGSM-Gpn010}.
It is observed that there is direct transition occurring at $h_{t} \approx 1.22$ between the zigzag phase and the paramagnetic phase.

\subsection{finite-size scaling of the magnetization}

\begin{figure}[!ht]
\centering
\includegraphics[width=0.75\columnwidth, clip]{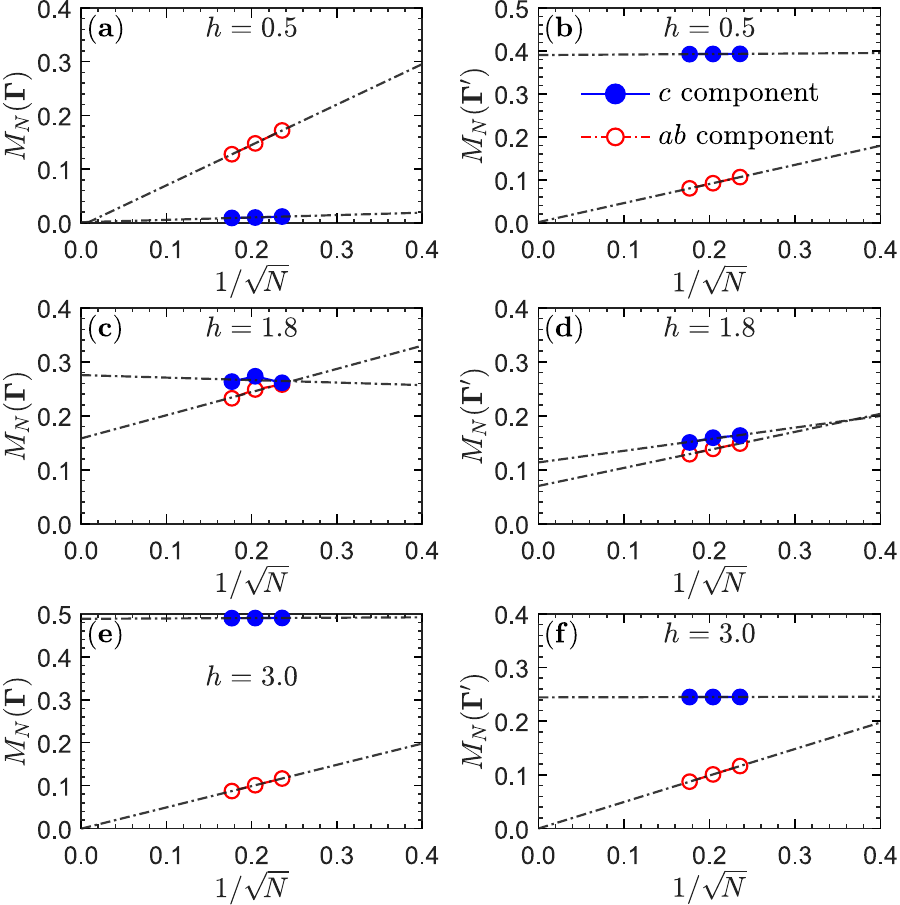}\\
\caption{Extrapolation of the magnetic order parameters $M_N(\boldsymbol{\Gamma})$ and $M_N(\boldsymbol{\Gamma}')$ as a function of magnetic field $h$
  in the $\Gamma$-$\Gamma'$ model with $\Gamma'/\Gamma$ = 0.3, with $N$ = 18, 24, and 32.
  In each panel, the blue filled circles stand for the values along the $c$ direction, while the red open circles represent the value in the $ab$ plane.
  The magnetic field is $h$ = 0.5 (a and b, AFM$_c$ phase), $h$ = 1.8 (c and d, spin-flop phase), and $h$ = 3.0 (e and f, paramagnetic phase).
  The error bars are much smaller than the size of the symbols.
  }\label{FIGSM-Gp030}
\end{figure}

We take the $\Gamma$-$\Gamma'$ model in a [111] magnetic field as an example to show the finite-size correlation and extrapolation.
Without lose of generality, we will fix $\Gamma'/\Gamma$ = 0.3 and choose magnetic field $h$ = 0.5 (AFM$_c$ phase), 1.8 (spin-flop phase), and 3.0 (paramagnetic phase).
The system size $N$ is chosen as 18, 24, and 32. The extrapolations of the magnetic order parameters $M_N(\boldsymbol{\Gamma})$ and $M_N(\boldsymbol{\Gamma}')$
as functions of $1/N$ are shown in Fig.~\ref{FIGSM-Gp030}(a)-(f).
In each panel, the blue filled circles stand for the values along the $\textbf{c} [111]$ direction,
while the red open circles represent the value in the $\textbf{a} [11\bar{2}]$-$\textbf{b} [\bar{1}10]$ plane.
\begin{itemize}
  \item In panel (a) and (b) [AFM$_c$ phase], we find that only the $c$ component (blue circles) of $M_N(\boldsymbol{\Gamma}')$ is finite, in accordance with the property of the AFM$_c$ phase.
  \item In panel (e) and (f) [paramagnetic phase], the $c$ components (blue circles) of $M_N(\boldsymbol{\Gamma})$ and $M_N(\boldsymbol{\Gamma}')$  are both finite. The difference lies in that, the value of $M_N(\boldsymbol{\Gamma})$ is around 0.5, while the value of $M_N(\boldsymbol{\Gamma'})$ is around 0.25.
  \item In panel (c) and (d) [spin-flop phase], we find that not only the $c$ components (blue circles) of $M_N(\boldsymbol{\Gamma})$ and $M_N(\boldsymbol{\Gamma}')$ are finite, but also the $ab$ components (red circles) are nonzero.
\end{itemize}

\begin{figure}[!ht]
\centering
\includegraphics[width=0.75\columnwidth, clip]{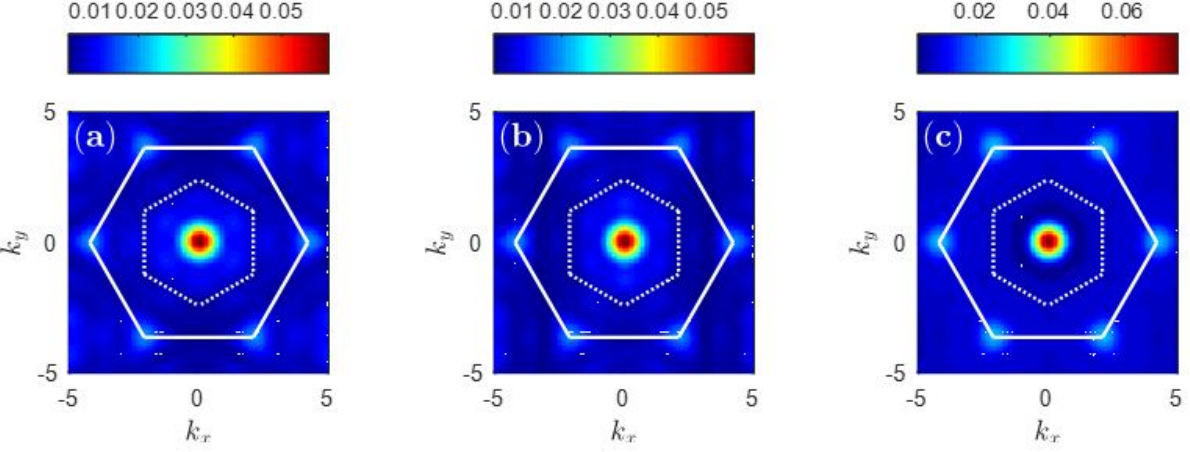}\\
\caption{The static structure factor of the spin-flop phase in the $\Gamma$-$\Gamma'$ model in the [111] magnetic field with ($\Gamma'/\Gamma$ = 0.3, $h/\Gamma$=1.8).
  Panels (a), (b), and (c) show the static structure factor along the $\mathbf{a}$ direction, $\mathbf{b}$ direction, and $\mathbf{c}$ direction, respectively.
  }\label{FIGSM-SpinFlopSSF}
\end{figure}

Finally, for the sake of clarity, we show the static structure factor of the spin-flop phase in Fig.~\ref{FIGSM-SpinFlopSSF}.
Since the spins in the spin-flop phase have a dominant component along the [111] direction, we now define the $\alpha$ ($= x,y,z$) component of the static structure factor as
\begin{equation}\label{SMEQ:SSF}
\tilde{\mathbb{S}}_{N}^{\alpha}({\bm{q}}) = \frac{1}{N}\sum_{ij} \langle \tilde{S}^{\alpha}_i \tilde{S}^{\alpha}_j\rangle e^{i{\bm{q}}\cdot{({\bm{R}}_i-{\bm{R}}_j)}}
\end{equation}
where
$\tilde{S}^{x}_i = \frac{1}{\sqrt6}(S_i^x + S_i^y - 2S_i^z)$,
$\tilde{S}^{y}_i = \frac{1}{\sqrt2}(S_i^y - S_i^x)$,
$\tilde{S}^{z}_i = \frac{1}{\sqrt3}(S_i^x + S_i^y + S_i^z)$.
Here, ${\bm{R}}_i$ is the position of site $i$ and $\bm{q}$ is the wavevector in the reciprocal space.
Panels (a), (b), and (c) represent, respectively, the structure factor along the
$\mathbf{a}\ \left[11\bar{2}\right]$ direction (i.e., $\tilde{\mathbb{S}}_{N}^{x}({\bm{q}})$),
$\mathbf{b}\ \left[\bar{1}10\right]$ direction (i.e., $\tilde{\mathbb{S}}_{N}^{y}({\bm{q}})$),
and $\mathbf{c}\ \left[111\right]$ direction (i.e., $\tilde{\mathbb{S}}_{N}^{z}({\bm{q}})$).
The difference between the spin-flop phase and the totally polarized phase is that the $a$ and $b$ components of the structure factor are finite and have an equal strength.

\section{Quantum $\Gamma$-$\Gamma'$ model: excitation gap under cylinder boundary condition}

\begin{figure}[!ht]
\centering
\includegraphics[width=0.45\columnwidth, clip]{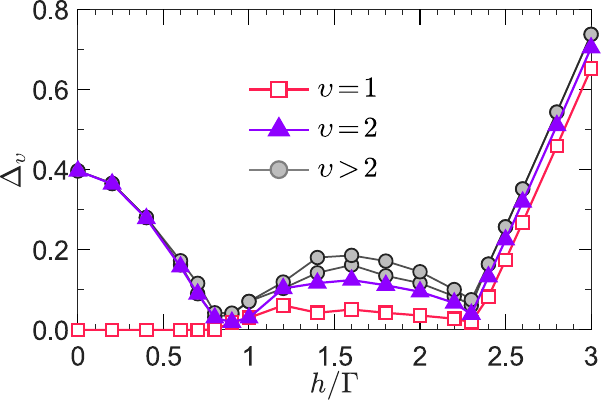}\\
\caption{The first four excitation gaps $\Delta_{\upsilon}$ ($\upsilon$ = 1--4) as a function of magnetic field $h/\Gamma$ in the $\Gamma$-$\Gamma'$ model
  with $\Gamma'/\Gamma = 0.3$. Here, the computation is executed on a $12\times6$ YC cylinder under cylinder boundary condition.
  }\label{FIGSM-Gap12x06}
\end{figure}

In Fig.~\textcolor{red}{10}(a) in the main text, we show the excitation gaps $\Delta_{\upsilon} = E_{\upsilon}-E_0$ obtained on a 24-site cluster.
Here, we present the first few excitation gaps $\Delta_{\upsilon}$ obtained on a $12\times6$ YC cylinder, see Fig.~\ref{FIGSM-Gap12x06}.
It is found that there are two quantum phase transitions occurring at $h_{t1} \approx 0.8$ and $h_{t2} \approx 2.3$.
The values of the transition points are quite close to the estimated results revealed in the 24-site cluster.
Such a consistency implies the robustness of the intermediate spin-flop phase.

% \bibliography{Supplementary}
%
%%%%%%%%%%%%%%%%%%%%%%%%%%%%%%%%%%%%%%%%%%%%%%%%%%%%%%%%%%%%%%%%%%%%%%%%%%%%%%%%%%%%%%%%%%%%%%%%%%%%%%%%%%%%%%%%%%%%%%%%%%%%%%%%%%%%%%%%%%%%%%%%

\end{document}